\documentclass[
 reprint,
 superscriptaddress,
 amsmath,amssymb,
]{revtex4-2}

\usepackage{graphicx}
\usepackage{dcolumn}
\usepackage{bm}

\usepackage{color}
\usepackage{braket}
\usepackage[thinlines]{easytable}

\begin{document}

\title{Scalable and Programmable Phononic Network with Trapped Ions}

\affiliation{
 State Key Laboratory of Low Dimensional Quantum Physics, Department of Physics, Tsinghua University, Beijing 100084, China
 }
\affiliation{
 QOLS, Blackett Laboratory, Imperial College London, London SW7 2AZ, United Kingdom
 }

\author{Wentao Chen}
 \email{
 chen-wt17@mails.tsinghua.edu.cn
 }
 \affiliation{
 State Key Laboratory of Low Dimensional Quantum Physics, Department of Physics, Tsinghua University, Beijing 100084, China
 }

\author{Yao Lu}
 \affiliation{
 State Key Laboratory of Low Dimensional Quantum Physics, Department of Physics, Tsinghua University, Beijing 100084, China
 }
 \affiliation{
 Shenzhen Institute for Quantum Science and Engineering, Southern University of Science and Technology, Shenzhen 518055, China
 }

\author{Shuaining Zhang}
 \affiliation{
 State Key Laboratory of Low Dimensional Quantum Physics, Department of Physics, Tsinghua University, Beijing 100084, China
 }
 \affiliation{
 Department of Physics, Renmin University of China, Beijing 100872, China
 }

\author{Kuan Zhang}
 \affiliation{
 State Key Laboratory of Low Dimensional Quantum Physics, Department of Physics, Tsinghua University, Beijing 100084, China
}
 \affiliation{
 MOE Key Laboratory of Fundamental Physical Quantities Measurement, Hubei Key Laboratory of Gravitation and Quantum Physics, PGMF, Institute for Quantum Science and Engineering, School of Physics, Huazhong University of Science and Technology, Wuhan 430074, China
}

\author{Guanhao Huang}
 \affiliation{
 State Key Laboratory of Low Dimensional Quantum Physics, Department of Physics, Tsinghua University, Beijing 100084, China
 }

\author{Mu Qiao}
 \affiliation{
 State Key Laboratory of Low Dimensional Quantum Physics, Department of Physics, Tsinghua University, Beijing 100084, China
 }
 
\author{Xiaolu Su}
 \affiliation{
 State Key Laboratory of Low Dimensional Quantum Physics, Department of Physics, Tsinghua University, Beijing 100084, China
 }
 
\author{Jialiang Zhang}
 \affiliation{
 State Key Laboratory of Low Dimensional Quantum Physics, Department of Physics, Tsinghua University, Beijing 100084, China
 }

\author{Jingning Zhang}
 \affiliation{
 Beijing Academy of Quantum Information Sciences, Beijing 100193, China
 }

\author{Leonardo Banchi}
 \affiliation{
 Dipartimento di Fisica e Astronomia, Universit\`a di Firenze, I-50019, Sesto Fiorentino (FI), Italy
 }
 \affiliation{
 INFN, Sezione di Firenze, I-50019, Sesto Fiorentino (FI), Italy
 }

\author{M.S. Kim}
 \email{
 m.kim@imperial.ac.uk
 }
 \affiliation{
 QOLS, Blackett Laboratory, Imperial College London, London SW7 2AZ, United Kingdom
 }

\author{Kihwan Kim}
 \email{
 kimkihwan@mail.tsinghua.edu.cn
 }
 \affiliation{
 State Key Laboratory of Low Dimensional Quantum Physics, Department of Physics, Tsinghua University, Beijing 100084, China
 }
 \affiliation{
 Beijing Academy of Quantum Information Sciences, Beijing 100193, China
 }
 \affiliation{
 Frontier Science Center for Quantum Information, Beijing 100084, People’s Republic of China
 }

\date{\today}

\begin{abstract}
Controllable bosonic systems can provide post-classical computational power with sub-universal quantum computational capability. A network that consists of a number of bosons evolving through beam-splitters and phase-shifters between different modes, has been proposed and applied to demonstrate quantum advantages ~\cite{aaronson2011computational,spring2013boson,broome2013photonic,tillmann2013experimental,crespi2013integrated,carolan2015universal,wang2017high,wang2019boson,zhong2020quantum,arrazola2021quantum,brod2019photonic}. While the network has been implemented mostly in optical systems with photons, 
recently alternative realizations have been explored, where major limitations in photonic systems such as photon loss, and probabilistic manipulation can be addressed~\cite{garcia2019simulating,qi2020regimes,quesada2020exact}.
Phonons, the quantized excitations of vibrational modes, of trapped ions can be a promising candidate to realize the bosonic network~\cite{Lau2012Proposal,Shen2014Scalable,chen2021quantum,brown2011coupled,harlander2011trapped,toyoda2015hong,debnath2018observation,fluhmann2019encoding,tamura2020quantum,nguyen2021experimental,Um16Phonon,shen2018quantum}. Here, we experimentally demonstrate a minimal-loss phononic network that can be programmed and in which any phononic states are deterministically prepared and detected. We realize the network with up to four collective-vibrational modes, which can be straightforwardly extended to reveal quantum advantage. We benchmark the performance of the network with an exemplary algorithm of tomography for arbitrary multi-mode states with a fixed total phonon number~\cite{banchi2018multiphoton}. We obtain reconstruction fidelities of 94.5$\pm 1.95 \%$ and 93.4$\pm 3.15 \%$ for single-phonon and two-phonon states, respectively. Our experiment demonstrates a clear and novel pathway to scale up a phononic network for various quantum information processing beyond the limitations of classical and other quantum systems.
\end{abstract}

\maketitle

\begin{figure*}
    \centering
    \includegraphics[width=1\linewidth]{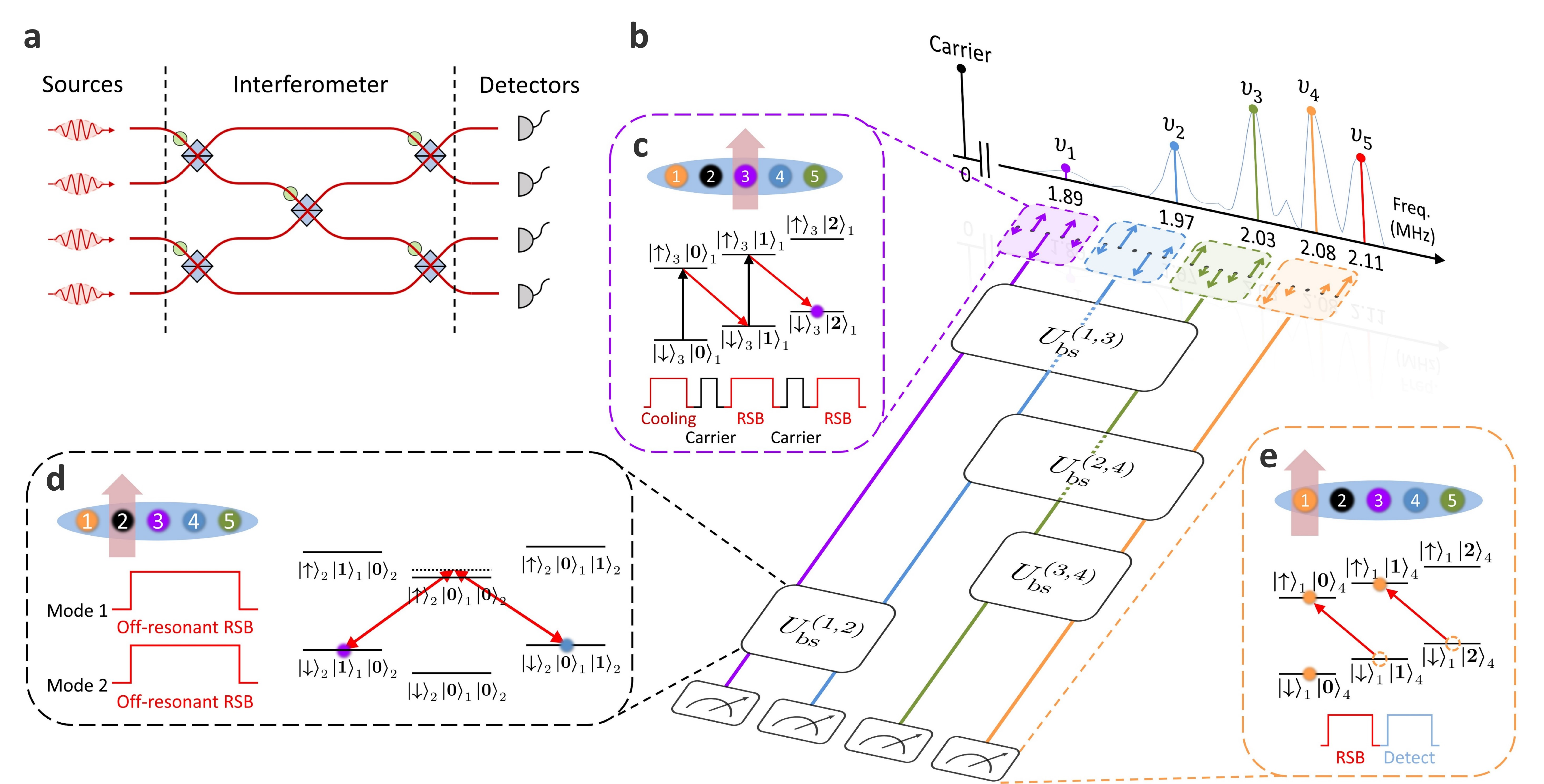}
    \caption{
    $\bm{\textbf{Overview of phononic network}}$. $\bm{\textbf{a}}$, Structure of a Bosonic network. The Bosonic network consists of three parts: input sources, interferometer, and detectors as illustrated by a linear optical system. The interferometer is composed of phase-shifters (green) and beam-splitters (blue). $\bm{\textbf{b}}$, A phononic network that consists of four vibrational modes denoted as $\nu_1$, $\nu_2$, $\nu_3$, and $\nu_4$. The phononic network also contains three parts: 
    $\bm{\textbf{c}}$, input-state preparation, 
    $\bm{\textbf{d}}$, programmable beam-splitting operations, and $\bm{\textbf{e}}$, detection. The upper part shows the experimentally measured mode spectrum, where the peaks are indicated by colored dots and lines. The mode vectors depending on ions are indicated by arrows inside dashed boxes with corresponding colors. 
    $\bm{\textbf{c}}$, State preparation. The preparation sequence of a Fock state of $\ket{n=2}$ for mode 1 is illustrated as an example, which is realized by applying twice of carrier ($\ket{\downarrow} \leftrightarrow \ket{\uparrow}$) and red-sideband ($\ket{\uparrow}\ket{n} \leftrightarrow \ket{\downarrow}\ket{n+1}$, where $n\ge 0$) transitions after vibrational ground state cooling. Raman laser beams are used to manipulate the internal and vibrational energy levels, serving as a global (blue) and an individually addressing (red) coupling for the full control of the phononic network. Here the ion 3 is chosen due to its largest coupling strength on mode 1 as shown in the mode vector (see Methods B). $\bm{\textbf{d}}$, Beam-splitting operation. Two vibrational modes are connected by applying two non-resonant RSB transitions with the same detuning on the same ion, which effectively realize a beam-splitting operation (see Methods C). As an example, ion 2 is chosen for operation between mode 1 and mode 2 because the product of coupling strengths is the largest. $\bm{\textbf{e}}$, State detection. States of vibrational modes are detected with assigned ions, where no fluorescence for zero phonon states and fluorescence for non-zero phonon states. Zero phonon state and non-zero phonon states are projected to qubit-down $\left|\downarrow \right\rangle$ (no fluorescence) and qubit-up states $\left|\uparrow \right\rangle$ (fluorescence), respectively, by applying uniform red-sideband transitions~\cite{an2015experimental,Um16Phonon} (see Methods for details). For mode 4, as an example, ion 1 is chosen for the operation due to its largest coupling strength. The designated ions for the preparation and detection of vibrational modes are marked in the same color as the modes.
    } 
    \label{fig:Figure_1}
\end{figure*}

\begin{figure}
	\centering
	\includegraphics[width=1\linewidth]{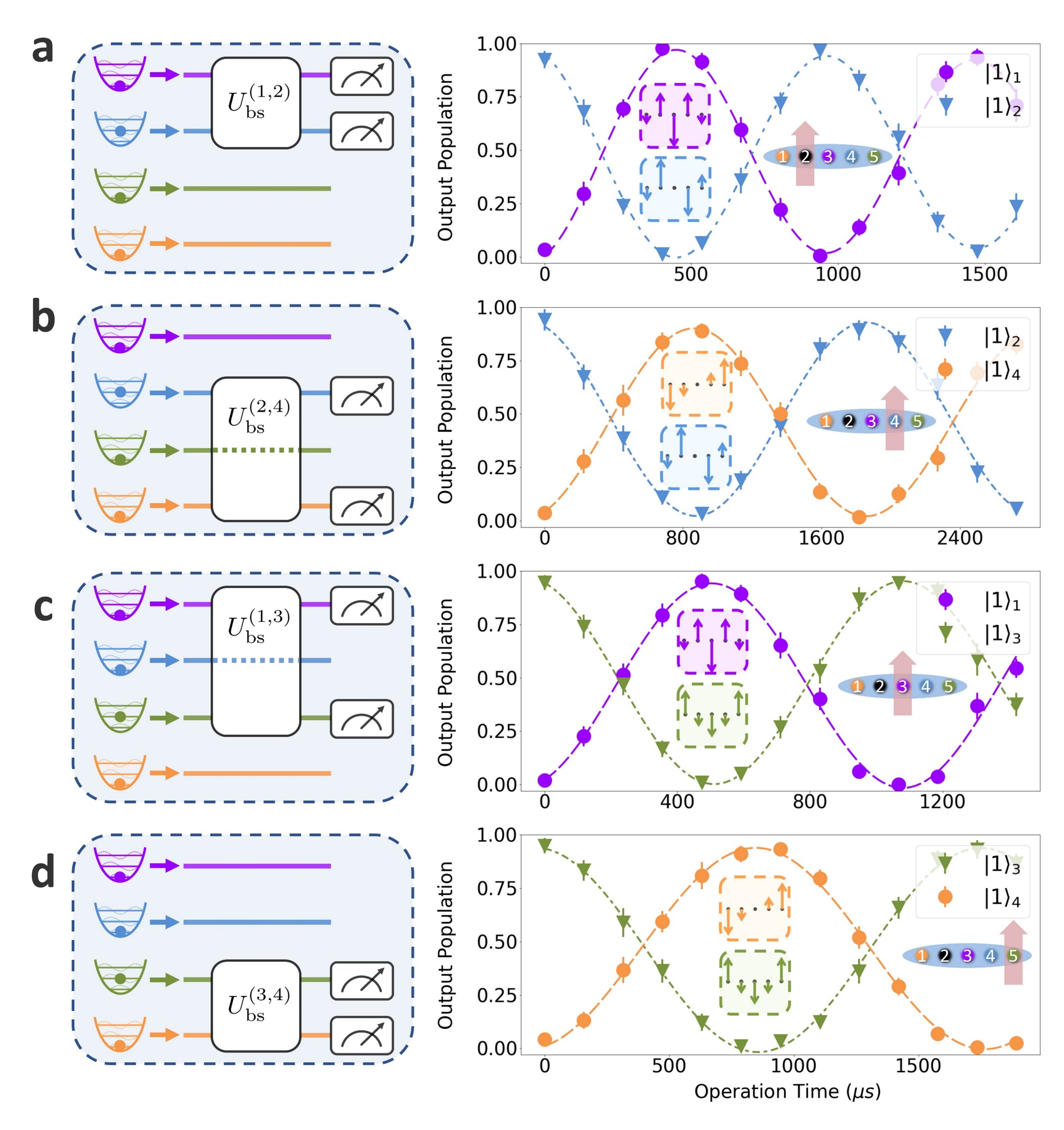}
	\caption{
	$\bm{\textbf{Beam-splitters}}$. Experimental results from time-evolution under the beam-splitter interactions for the different modes: \textbf{a}, between mode 1 and mode 2; \textbf{b}, between mode 2 and mode 4; \textbf{c}, between mode 1 and mode 3; \textbf{d}, between mode 3 and mode 4. The left panel shows the quantum circuits of each beam-splitter. We prepare an initial state as shown in the beginning of the circuit, and perform a beam-splitter operation between the chosen modes. The right panel shows the population distributions of the phonon states in the chosen modes, with dots denoting the experiment results, and dashed lines the fitting results of the population of each mode. The ion used for the beam-splitter is marked inside the right figure. $\left|n \right\rangle_m$ shows the number of phonons $n$ in the $m$-th mode and zero phonon states are omitted. All the error bars occur in the figures represent 95$\%$ confidence intervals.
	}
	\label{fig:Figure_2}
\end{figure}

\begin{figure*}
	\centering
	\includegraphics[width=1\linewidth]{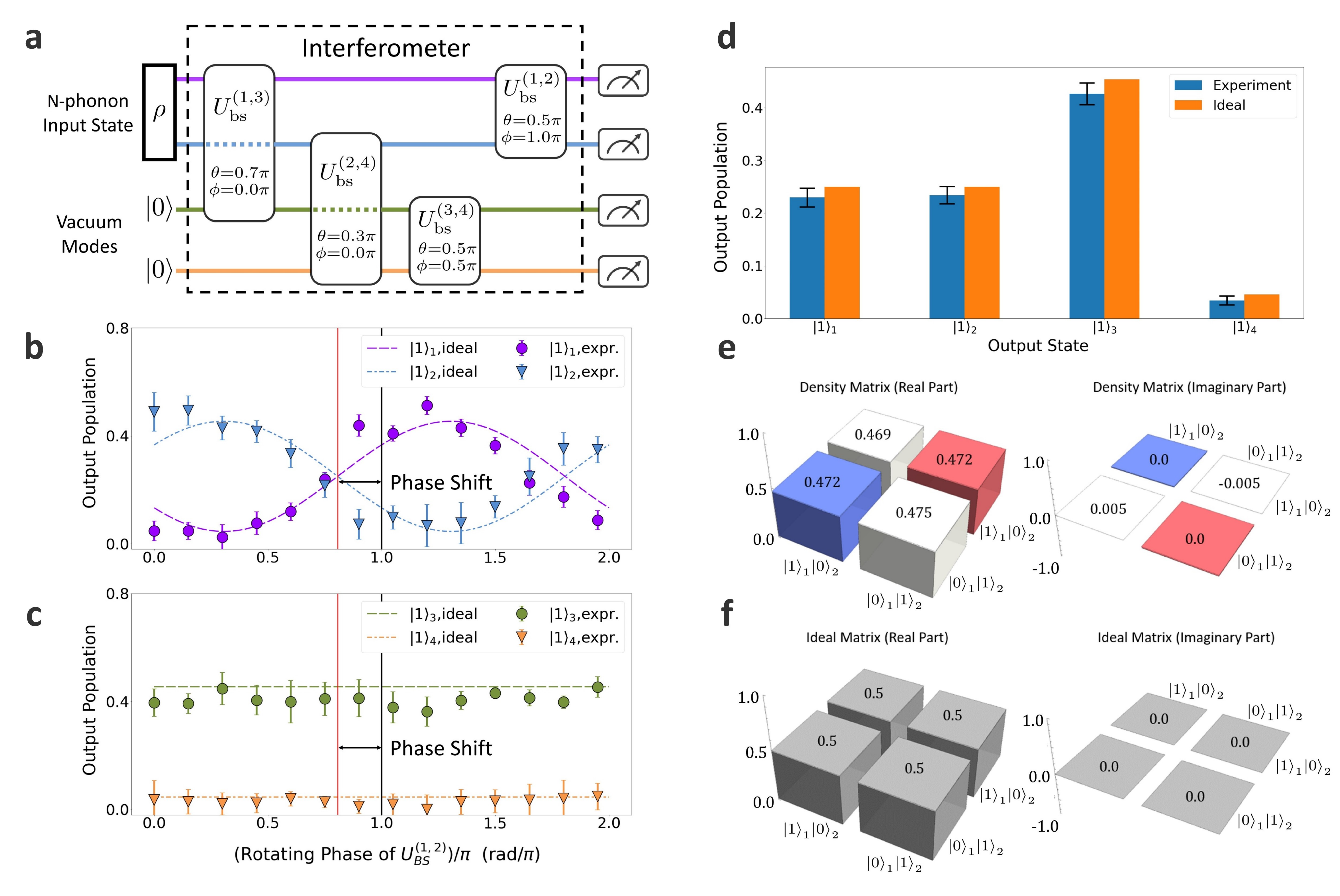}
	\caption{
    $\bm{\textbf{Tomography experiment with single phonon}}$.
	$\bm{\textbf{a}}$, Four-mode interferometric configuration for tomography of two-mode input states. Two ancillary vacuum states are required for reconstruction with a single configuration. Here $\theta$ and $\phi$ denote respectively for the optimal choice of rotating angle and phase of each beam-splitter for the tomography.
	Output population of states $\bm{\textbf{b}}$, $\left|1 \right\rangle_1$ and $\left|1 \right\rangle_2$ and $\bm{\textbf{c}}$, $\left|1 \right\rangle_3$ and $\left|1 \right\rangle_4$, depending on a phase of beam-spiltter $U^{(1,2)}_{\text{bs}}$ in the interferometer. The experiment outputs of phonon states (points) are compared with theoretically predicted values without any fitting parameters (dashed lines). The red line denotes for the experimentally chosen value for the interferometer. It is shifted from the ideal value ($\pi$) due to ac Stark shifts induced by previous beam-splitters, which are in agreement with calculations.
	$\bm{\textbf{d}}$, Output population of phonon states after interferometric configurations with programmed parameters shown in Fig.3a. The blue bars denote the experiment outputs and orange bars for the ideal values.
	$\bm{\textbf{e}}$, Reconstructed density matrix from experiment results.
	$\bm{\textbf{f}}$, Density matrix of the ideal input state. All the error bars occur in the figures represent 95$\%$ confidence intervals.
	}
	\label{fig:Figure_3}
\end{figure*}

\begin{figure}[!htb]
	\centering
	\includegraphics[width=1\linewidth]{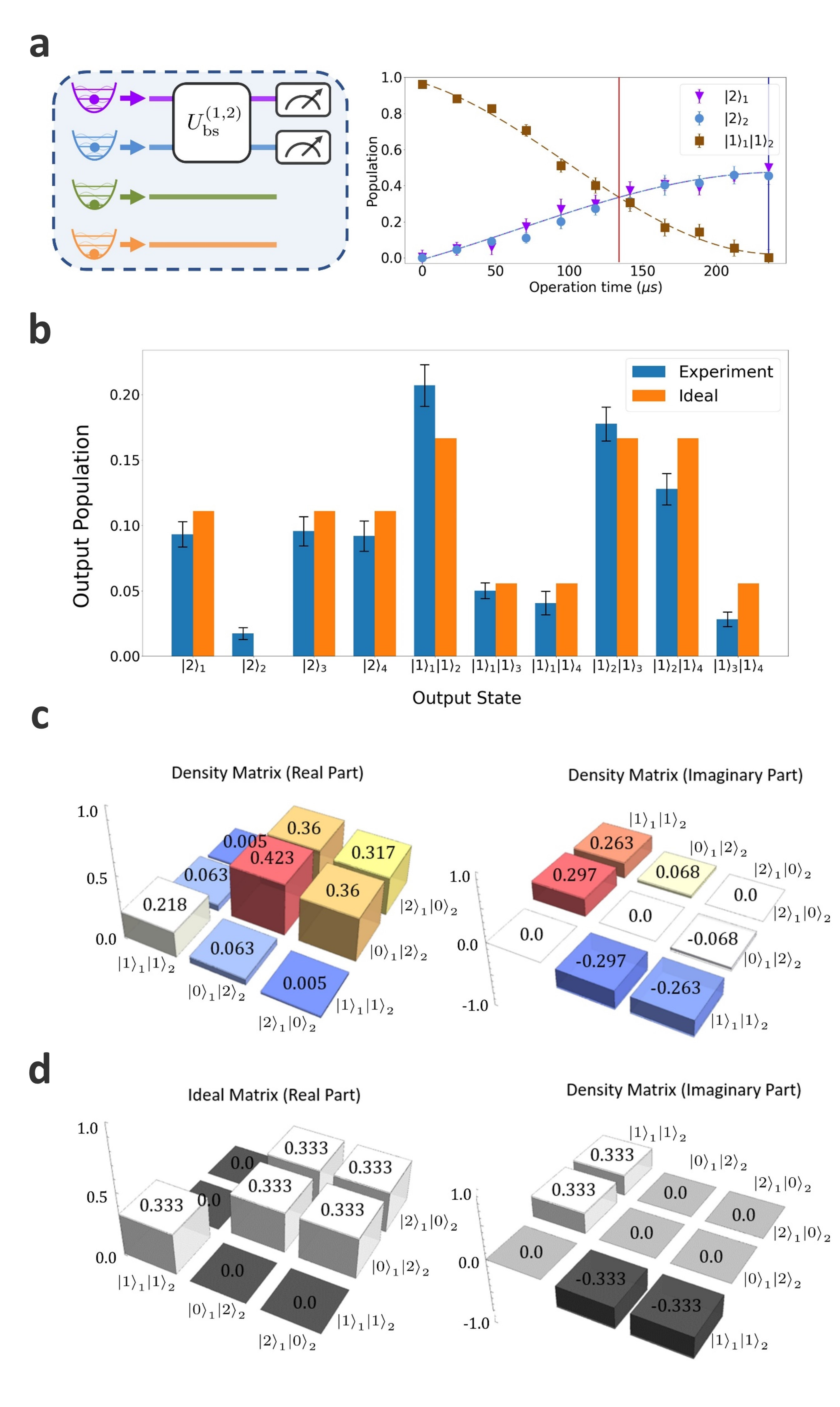}
	\caption{
	$\bm{\textbf{Tomography experiment with two phonons}}$.
	$\bm{\textbf{a}}$, Two-phonon experimental results from time-evolution under beam-splitter between mode 1 and mode 2, which is used for preparing the initial state $\frac{1}{\sqrt{3}}(|1\rangle_{1}|1\rangle_{2}+i|0\rangle_{1}|2\rangle_{2}+i|2\rangle_{1}|0\rangle_{2})$. Here the dots denoting the experiment results, dashed lines are theoretical curves. The red vertical line is the chosen point for preparing the initial state, and blue vertical line shows Hong-Ou-Mandel dip with a visibility of 99.7$\%$ and depth of 0.003. $\bm{\textbf{b}}$, Output population distributions of a two-phonon input state after interferometric configurations shown in Fig.3a. A total of 10 possible output states are detected. The blue bars denote the experiment outputs, and the orange bars show the ideal values. $\bm{\textbf{c}}$, Reconstructed density matrix from experiment results. $\bm{\textbf{d}}$, Ideal density matrix of the input state. All the error bars occur in the figures represent 95$\%$ confidence intervals.
	}
	\label{fig:Figure_4}
\end{figure}

\section{\label{sec:1}Introduction}

It is of interest to demonstrate the power of quantum computers that outperform their classical counterparts for certain problems~\cite{aaronson2011computational}. Bosonic systems spanning a large Hilbert spaces offer promising and useful applications. For instance, a network composed of a number of bosons evolving through beam-splitters and phase-shifters between different modes, has been proposed and applied to demonstrate quantum advantages~\cite{spring2013boson,broome2013photonic,tillmann2013experimental,crespi2013integrated,carolan2015universal,wang2017high,wang2019boson,zhong2020quantum,arrazola2021quantum}. Boson sampling devices can also be applied to solve quantum chemistry problems~\cite{huh2015boson,sawaya2019quantum,shen2018quantum,banchi2020molecular}, enhance stochastic algorithms~\cite{arrazola2018using,arrazola2018quantum,bradler2021graph} or quantum machine learning~\cite{schuld2019quantum,chabaud2021quantum}. The network has been implemented mostly in optical systems with photons~\cite{brod2019photonic}. However, technical bottlenecks exist in photon systems. In particular, photon loss and non-deterministic generation and inefficient detection of photonic states hinder their further scalability and demonstration of quantum advantages~\cite{garcia2019simulating,qi2020regimes,quesada2020exact}. It is thus desirable to explore new experimental platforms.  

In a trapped ion system, the quantized vibrational modes give rise to phonons that can be used as alternative bosons to build a bosonic network~\cite{Lau2012Proposal,Shen2014Scalable,chen2021quantum}. 
The phonon states can be deterministically prepared and detected by coupling between internal states of ions and the vibrational modes~\cite{leibfried2003quantum}. Recently, there have been various developments for phononic networks, but building programmable scalable networks remains a challenge. In principle, the number of vibrational modes can be increased by simply adding more ions in the system, which can be divided into two categories: local modes and collective modes. When ions are confined in separated trap-potentials or the distances between ions in a single trap-potential are relatively large, the vibration of an ion is localized and almost independent of the motion of the other ions. This is the local phonon mode regime~\cite{brown2011coupled,harlander2011trapped,toyoda2015hong,debnath2018observation,fluhmann2019encoding}. With two local modes, the coupling and hopping in the level of a single quanta~\cite{brown2011coupled,harlander2011trapped}, and Hong-Ou-Mandel interference~\cite{toyoda2015hong} have been observed. The hopping of a single phonon was extended to four local modes~\cite{tamura2020quantum} and a blockade of a single phonon in three local modes has also been demonstrated~\cite{debnath2018observation}. However, the coupling between local modes is always present due to the Coulomb interaction of ions, which makes it challenging to initialize and detect phonon states with high fidelities, or to perform the desired operations~\cite{debnath2018observation,ohira2019phonon}. In addition, local phonons are susceptible to electric field noise, which cause undesirable heating that degrades phonon-number conservation and the performance of the operations~\cite{brown2011coupled,harlander2011trapped}. 

In contrast, collective modes are considered when ions are tightly confined in the trap~\cite{James98}. The phonon number can be conserved in most of the collective modes, since they are not susceptible to homogeneous electric field noise~\cite{brownnutt2015ion,kalincev2021motional}. The coupling between collective modes is present only when laser beams with properly adjusted parameters on ions~\cite{shen2018quantum,Zhang2018Experimental}. Advanced controls with the collective vibrational modes have been demonstrated to generate NOON states and conditional operations~\cite{Zhang2018Experimental,Maslennikov2019quantum,Gan2020hybrid,nguyen2021experimental}. However, earlier experimental realizations were limited to single- and two-ion systems and it has not been shown how to scale up the system to our best knowledge. In this work, with up to four collective modes from five ions, we demonstrate the phononic network that contains all the essential operations including deterministic preparation and detection, and programmable beam-splitters, which are performed in a scalable manner. The phonon states in each mode are deterministically prepared and detected using Raman transition on an ion that couples the mode with large strength~\cite{leibfried2003quantum}. Beam-splitters between any two different modes are also implemented by applying Raman transition on an ion that strongly couples both modes~\cite{marshall2017linear,shen2018quantum}, where the rotating angle and phase of the beam-splitter can be controlled by the duration and the phase of the Raman beams, respectively.

\section{\label{sec:2}Preparation and Detection of Phonon States}

Our phononic network consists of the four collective vibrational modes in one of the radial directions, except for the center of mass mode of a five ion linear chain, as a result of the Coulomb interaction between the ions. The frequencies of modes are $\{ \nu_1, \nu_2, \nu_3, \nu_4, \nu_5 \}=2\pi\times\{1.905, 1.985, 2.057, 2.114, 2.153 \}$ MHz. As shown in Fig.~\ref{fig:Figure_1}a, a phononic network consists of three main procedures: (i) state preparation, (ii) a programmbale interferometer and (iii) output measurements. These procedures are realized by using individually addressed Raman-laser beams that manipulate the interaction between ion-qubits and vibrational modes. The ion-qubits are encoded in hyperfine levels of $^{171}$Yb$^+$ ions in $^2 S_{1/2}$ manifold denoted as $\left|\downarrow\right\rangle \equiv \left| F=0, m_F=0 \right\rangle$ and $\left|\uparrow\right\rangle \equiv \left| F=1, m_F=0 \right\rangle$ with an energy splitting of $\omega_0 = 12.642812$ GHz. The ion-qubit is initialized to $\left|\downarrow\right\rangle$ by the standard optical pumping method and measured through a qubit-state dependent fluorescence, which is individually detected by a multi-channel photomultiplier tube (PMT) (see Method A).

We deterministically prepare and detect phonon states of the vibrational modes of interest by using a properly chosen single ion (see Method B). We choose the ion with the largest coupling strength for the manipulation of a mode. In Fig.~\ref{fig:Figure_1}c, we illustrate the preparation of two-phonon state of mode 1 ($\ket{n=2}_1$) by using ion 3, as an example. A zero-phonon state ($\ket{\downarrow}_3\ket{0}_1$) in Fig.~\ref{fig:Figure_1}c is prepared by Doppler cooling (1 ms) and resolved sideband cooling (3 ms)~\cite{leibfried2003quantum}. Then the initial state of two-phonon state ($\ket{\downarrow}_3\ket{2}_1$) is prepared with a combination of carriers (single qubit rotations) and red-sideband transitions (RSB) individually applied on ion 3 as shown in Fig.~\ref{fig:Figure_1}c. As shown in Fig.~\ref{fig:Figure_1}e, the detection of the phonon state of the 4th mode is realized by an adiabatic RSB transition~\cite{an2015experimental,Um16Phonon,lv2017reconstruction,chen2021quantum} followed by qubit-state detection sequence, where no fluorescence for only $\left| n=0 \right\rangle$ and fluorescence for any state of $\left| n>0 \right\rangle$. To minimize the operation time, all the adiabatic RSB's are applied simultaneously on the chosen ions. The duration of carrier-$\pi$-pulse is 3 $\mu$s and that of red-sidebands is about 200 $\mu$s.

We realize a programmable interferometer that consists of beam-splitters with arbitrary rotating angle ($\theta_{\rm bs}$) and phase ($\phi_{\rm bs}$) by Raman transitions on a single ion (see Method C). Benefiting from full connectivity between collective vibrational modes and ions, a beam-splitter between arbitrary pairs of modes can be realized. Two Raman transitions which have the same frequency detuning $\Delta_{\rm bs}$ from RSB of two modes (labeled with $m$ and $n$) are applied to the single ion (labeled with $j$) at the same time, with different Rabi frequencies $\Omega_{j,m}$ and $\Omega_{j,n}$ which satisfies $\eta_{j,m} \Omega_{j,m} = \eta_{j,n} \Omega_{j,n} \approx \Delta_{\rm bs}/2$. The evolution operator of the effective Hamiltonian takes the form,
\begin{equation}
    \label{eq:one}
	U_{\text{bs},j}^{(m,n)}(t)=\exp[{i \theta_{\text{bs}}(t) \sigma_{z,j}(a_m a_n^{\dagger}e^{-i\phi_{\text{bs}}}+a_m^{\dagger}a_n e^{i\phi_{\text{bs}}})}]
\end{equation}

\begin{equation}
	\theta_{\text{bs}}=\frac{\eta_{j,m} \Omega_{j,m} \eta_{j,n} \Omega_{j,n} t}{4\Delta_{\text{bs}}},
\end{equation}
where $a_m^{\dagger}$ ($a_m$) is the creation (annihilation) operator of the m-th mode. Here $\phi_{\text{bs}}=\phi_m-\phi_n$ is the phase difference between two Raman transitions, which can be considered as a phase-shifter integrated into beam-splitter. With this laser-activated beam-splitter, the mixing of different modes is no longer limited to the nearest neighbor and can be programmed arbitrarily to construct interferometers for different applications.

Fig.~\ref{fig:Figure_2} shows the performance of the beam-splitters between various pairs of modes. In the beginning, a single phonon state of each mode is prepared and detected by the schemes shown in Fig.~\ref{fig:Figure_1}c and~\ref{fig:Figure_1}e, respectively. The average fidelities of state preparation and detection for single- and two-phonon states are 96.7$\%$ and 95.6$\%$, respectively (see Method B). The typical durations of the state preparation and detection are about 300 and 200 $\mu$s, respectively. The imperfections mainly come from the intensity fluctuations of Raman laser beams and off-resonant couplings to other modes, which can be improved by further technological developments (see Method B).

With the beam-splitters of Eq.~\ref{eq:one} after initial state preparation, the phonon states are coherently evolved between two modes. The beam-splitter is realized through the ion with the largest mode-coupling strengths $\eta_{j,m}$ for the related pairs of modes (see Method C). For example, the ion 2 is used for the beam-splitters between mode 1 and mode 2. In Fig.~\ref{fig:Figure_2}, each data point is obtained by averaging over 300 repetitions. We fit the data using exponentially decaying sinusoids and obtained time constants over 10 ms, which is more than ten times longer than the duration of the 50:50 beamsplitter. The average population fidelity of the 50:50 beam-splitters is 95.6$\pm$1.72$\%$ including the errors by preparations and measurements. The fidelity of 50:50 beam-splitter itself is 99.1 $\%$, which is estimated by fitting the fidelity decay of multiple beam splitters. The fidelity of the beamsplitter can be further improved by suppressing heating and decoherence (see Method C).

\section{\label{sec:3}Tomography with a Programmable Phonon Network}
The performance of our phononic network is verified by demonstrating the boson sampling tomography protocol~\cite{banchi2018multiphoton}, which allows for the reconstruction of an arbitrary input state, with a definite total number of phonons in multiple modes, from measuring outcomes of the interferometric configurations. The number of configurations can be reduced to one when we include additional vacuum modes~\cite{banchi2018multiphoton}. Having access to the full tomography from the sampling data, we can easily verify and quantify the performance of our phononic network, in contrast to other sampling algorithms. 

In our realization, two vibrational modes are used for input states and the other two modes serve as ancillary vacuum modes. We choose modes 1 and 2 as input and modes 3 and 4 as ancillary modes. For a given input and interferometric configuration, the output probability of a basis state $\left|\nu^{\prime}\right\rangle$ of four modes takes the form as, 
\begin{eqnarray}
p_{\nu^{\prime}}=\left\langle\nu^{\prime}\left| U_{\textrm{IFO}}^{\dagger} \rho^{\prime} U_{\textrm{IFO}}\right| \nu^{\prime}\right\rangle
\label{eq:three},
\end{eqnarray}
where $\rho^{\prime} = \rho \otimes (\left| 0,0 \rangle \langle 0,0 \right|) $ denotes the density matrix of a four-mode state including two input modes and two ancillary vacuum modes, and $U_{\textrm{IFO}}$ is the unitary operation for the interferometric configuration. The reconstruction of the input state can be realized with a singe $U_{\textrm{IFO}}$ by measuring the probabilities $p_{\nu^\prime}$ for all possible output states.
 
Fig.~\ref{fig:Figure_3}a shows the interferometric configuration with optimal rotating angles and phases of four sequentially-applied-beam-splitters for efficient and reliable reconstruction of the density matrix (see Methods E). For a single-phonon case, we choose the input state $\left| \psi \right\rangle = (\left| 1 \right\rangle_1 \left| 0 \right\rangle_2+ \left| 0 \right\rangle_1\left| 1 \right\rangle_2)/\sqrt{2}$ as an example. We verify the phase coherence of our interferometer by scanning the phase $\phi_{\text{bs}}^{(1,2)}$ of the last beam-splitter $U_{\text{bs}}^{(1,2)}$. 
As shown in Fig.~\ref{fig:Figure_3}b, the changes of the final state populations are in agreement with the theoretical predictions without any fitting parameters, which clearly shows the programming capability for the parameters of beam-splitters (see Method D). 
Fig.~\ref{fig:Figure_3}c shows no influence of the $U_{\text{bs}}^{(1,2)}$ on unrelated modes 3 and 4, which implies negligible crosstalk of the beam-splitters. As shown in Fig.~\ref{fig:Figure_3}d, the final output populations for the interferometric setting of Fig.~\ref{fig:Figure_3}a are in agreement with the ideal values, which are used for the reconstruction of the density matrix of the input state. Only four output-states are detected because the input is a single-phonon state. Fig.~\ref{fig:Figure_3}e shows the reconstructed density matrix from the experimental results with a fidelity of $94.5 \pm 1.95\%$ and purity of $0.893$, in comparison to the ideal case shown in Fig.~\ref{fig:Figure_3}f. The fidelity includes operations for state preparation, state manipulation and measurements, which therefore shows the accuracy of our highly controllable platform.

The tomography of two-phonon state is also demonstrated as shown in Fig.~\ref{fig:Figure_4}. The two-phonon experiments contain pure quantum interference, i.e., Hong-Ou-Mandel interference as shown in Fig.~\ref{fig:Figure_4}a, which demonstrates the bosonic nature of the phononic network~\cite{toyoda2015hong,shen2018quantum}. We prepare an input state of $\left| \psi \right\rangle = (\left| 1 \right\rangle_1 \left| 1 \right\rangle_2+ i\left| 0 \right\rangle_1\left| 2 \right\rangle_2+ i\left| 2 \right\rangle_1\left| 0 \right\rangle_2)/\sqrt{3}$. With the same interferometric configuration of Fig.~\ref{fig:Figure_3}a, the output probabilities of the phononic network are shown in Fig.~\ref{fig:Figure_4}b, which are used for the reconstruction of the density matrix of the input state. The number of output states increase to ten with a larger phonon number input state. We note that the detections do not resolve phonon numbers, we assume the phonon numbers are conserved for the measurements of ten output-states (see Method B for details). The reconstructed density matrix from the experimental results is shown in Fig.~\ref{fig:Figure_4}c with a fidelity of $93.4 \pm 3.15\%$ and purity of $0.920$ in comparison to the ideal one shown in Fig.~\ref{fig:Figure_4}d. We do not observe any noticeable reduction of the fidelity in two-phonon experiment, which demonstrates high-quality performance of our platform with two phonons.

\section{\label{sec:4}Conclusion and Outlook}
Our programmable number-conserving phononic network of collective-transverse-vibrational modes can be scaled up to reach quantum advantages with large numbers of modes, $M$ and phonons, $N$. In our experiment, the average fidelity of preparing and detecting a single phonon in a mode is 96.70$\pm$1.31$\%$. When the number of phonons increases to $N$, the probability of detecting all the phonons scales as 0.967$^N$, which surpasses the success probability of photonic systems which scale as 0.3$^N$ for the best performance at the moment \cite{wang2017high,wang2019boson}. In our realization, the imperfection of each 50:50 beam splitter is around 1 $\%$ and a single phonon may pass through the number of $M-1$ beam splitters. In principle, we can simultaneously perform many beam splitters to directly create an arbitrary interferometer with the capability of full connectivity of our phononic network, which further suppresses imperfections \cite{lu2019global}.  

We estimate that about a hundred vibrational modes can be utilized for the phononic network in a single trap (see Method F). The number of phonons can be easily increased thanks to the deterministic preparation of phononic states. The duration of preparing phonon states for each mode increases with scaling of $\approx N_I$, where $N_I$ is the number of ions, due to the reduced Lamb-Dicke parameters. The number of phonons in each mode can be simultaneously prepared using properly assigned ions to the modes. More than single phonons at each mode can be deterministically prepared with the time cost scaling as  $\sum (1/\sqrt{n_m})$, where $n_m$ is the number of phonons in the mode $m$. Since the number of possible states in the bosonic network grows as $(N+M-1)!/N!/M!$, the increase of total phonon number $N$ for a given number of modes $M$ can be considered as a complementary implementation for a bosonic network to demonstrate quantum advantage. When the number of phonons is larger than the number of modes, it may be necessary to equip with number-resolving detections of phonons to obtain the full statistics of phonon distributions. Schemes capable of number-resolving detections have been realized in single mode and two modes~\cite{an2015experimental,Um16Phonon,shen2018quantum,Zhang2018Experimental} and can be further extended to any number of modes either with fast detection~\cite{noek2013high} or mapping to multi-levels of ions~\cite{ohira2019phonon}.

The phononic network can be extended for more complex phonon problems, such as a Gaussian Boson sampling. The various Gaussian states including coherent and squeezed states have been implemented in trapped ion systems. Moreover, the combination with the qubit-degrees of freedom in a trapped ion system can realize hybrid quantum computing with both discrete and continuous variables~\cite{Gan2020hybrid}. It may further enhance the capability of the phononic system, which can introduce nonlinearities on the vibrational modes \cite{ding2017cross} for applications to continuous-variable quantum computations \cite{fluhmann2019encoding,Gan2020hybrid} and quantum chemistry~\cite{huh2015boson,sawaya2019quantum,shen2018quantum,banchi2020molecular}.

\bibliography{Phonon}

\section*{Acknowledgements}
This work was supported by the National Key Research and Development Program of China under Grants No.\ 2016YFA0301900 and No.\ 2016YFA0301901, the National Natural Science Foundation of China Grants No.\ 92065205, and No.\ 11974200. MSK’s work was supported by the UK Hub in Quantum Computing and Simulation, part of the UK National Quantum Technologies Programme with funding from UKRI EPSRC grant EP/T001062/1 and by the Korea Institute of Science and Technology (KIST) Open Research Program. L.B. acknowledges support by the program “Rita Levi Montalcini” for young researchers. We thank Li You for carefully reading of the manuscript.

\section*{Author information}
\subsection*{Author contributions}
W.C., Y.L., S.Z, and K.Z. with assistance of X.S. and J.Z. developed the experimental system. L.B., and M.S.K. provided the theoretical idea and W.C., Y.L., and J.-N.Z. with help of G.H. and Q.M. optimized experimental schemes. W.C. took and analyzed the data. K.K. supervised the project. W.C., L.B., M.S.K. and K.K. contributed to the writing of the manuscript with the agreement of all the other authors.
\subsection*{Corresponding author}
Correspondence to W.C, M.S.K. and K.K.
\subsection*{Competing interests}
The authors declare no competing interests.
\section*{Data Availability}
All relevant data are available from the corresponding authors upon request.

\section*{Method} 

\subsection{Individual fluorescence detection of multiple ion-qubits} 

\begin{figure}[!htb]
    \centering
    \includegraphics[width=1\linewidth]{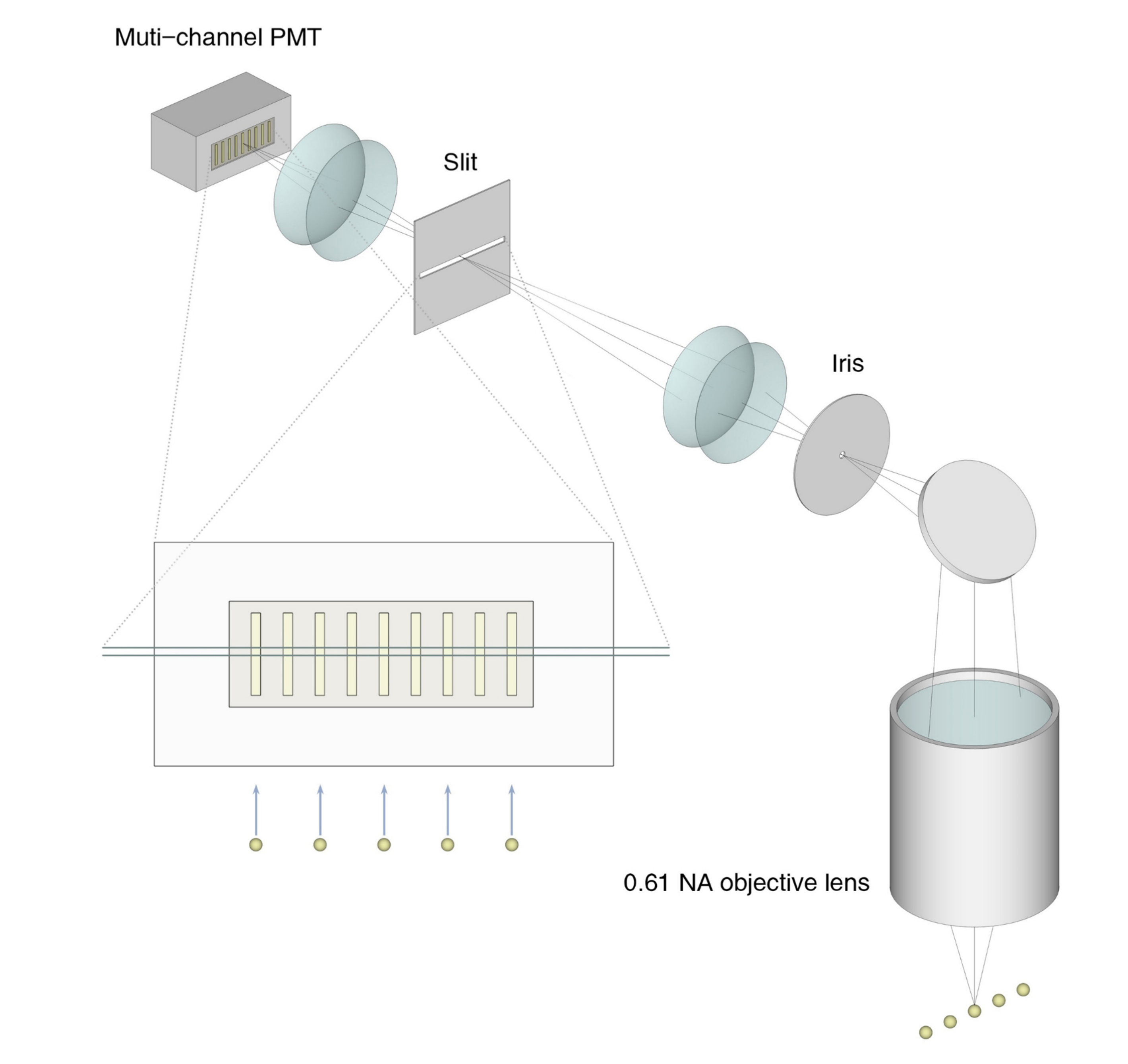}
    \caption{
    \textbf{Imaging system for the fluorescence detection of individual ion-qubit states.} The fluorescence of five ions is collected by a high NA lens and imaged to 32-channel PMT. To reduce crosstalk, we image each ion into alternative channels of PMT and put a slit at the second focused plane of the imaging system to suppress horizontal and vertical components, respectively.
    } 
    \label{fig:Figure_A1}
\end{figure}

Fig.~\ref{fig:Figure_A1} shows our scheme of ion-qubits detection. We use a high-NA objective lens (Photon Gear 15470-S) to collect the fluorescence from the five ions, then a 32-channel PMT (Hamamatsu H12211-01) detects the fluorescence of each ion. The average detection fidelity for all the five-qubit states is $(96.7\%)^5 \approx 84.6\%$ with a duration of 250 $\mu s$ and an average bright count of around 8. We estimate that for each ion, a $1.3\%$ error comes from the crosstalk and a $2\%$ error comes from the off-resonant optical pumping and the overlap of the photon state distribution between bright and dark states. To eliminate errors in detection, we utilize the detection-error correction method proposed in~\cite{shen2012correcting}. 

\subsection{State preparation and detection and heating-rate measurements}
\subsubsection{Adiabatic phonon state preparation and detection}
We use adiabatic sideband transitions to compensate for the Rabi frequencies difference between phonon number states to prepare and detect phonon states~\cite{an2015experimental,Um16Phonon,lv2017reconstruction}. The typical duration for the adiabatic sideband transition is 200 $\mu s$, three times larger than the $\pi$ time of ground-state sideband transition. We verify the performance of the adiabatic sideband transition by preparing and detecting the target states (one-phonon and two-phonon states) on each mode, the measurement results are shown in Table.~\ref{tab:SPD errors}.

\subsubsection{Heating-rate measurements}
Heating measurements of COM mode and other transverse modes are shown in Fig.~\ref{fig:Figure_B1} and~\ref{fig:Figure_B2}, respectively. Average phonon numbers after different waiting times from initial states can be achieved by fitting the measured results to the time evolution of thermal state distributions, which take the form of
\begin{equation}
\rho_{\mathrm{m}}(\bar{n})=\sum_{n} \frac{\bar{n}^{n}}{(\bar{n}+1)^{n+1}}|n\rangle_{m}\left\langle\left. n\right|_{m}\right.,
\end{equation}
where $\rho_{\mathrm{m}}$ denotes the density matrix of a thermal state on mode $m$, and $\bar{n}$ the average phonon number. We estimate a heating rate of $2.3 \times 10^3$ quanta/s on the COM mode as shown in Fig.~\ref{fig:Figure_B1}. For all the other four modes as shown in Fig.~\ref{fig:Figure_B2}, we do not observe the heating rates larger than error bars. We observe the average phonon number of 0.2 $\pm 0.3$ on modes 3 and 4, which can be corresponding to the rate of about 30 $\pm 45$ quanta/s. For modes 1 and 2, almost no heating effects are observed.

\begin{table}[!htb]
\caption{\label{tab:SPD errors}%
\textbf{Detection and preparation fidelities of each mode}. We prepared single-phonon states and two-phonon states in each mode and used adiabatic sideband transitions to perform projection measurements of the phonon states. We use the bright state population of the chosen ion after adiabatic sideband transitions as the population fidelity of state preparation and detection.
}
\begin{ruledtabular}
\begin{tabular}{ccc}
\textrm{Mode}&
\textrm{One-phonon State}&
\textrm{Two-phonon State}\\[0.3ex]
\colrule
1 & 96.52$\pm$1.81$\%$ & 96.12$\pm$1.23$\%$\\[0.3ex] 
2 & 97.46$\pm$1.19$\%$ & 95.03$\pm$1.89$\%$\\[0.3ex] 
3 & 96.66$\pm$1.05$\%$ & 96.03$\pm$1.64$\%$\\[0.3ex] 
4 & 96.16$\pm$1.20$\%$ & 95.25$\pm$1.49$\%$\\[0.3ex] 
Average & 96.70$\pm$1.34$\%$ & 95.61$\pm$1.58$\%$\\
\end{tabular}
\end{ruledtabular}
\end{table}

\subsection{Beam splitters} 
\subsubsection{Scheme of beam-splitters}
\begin{figure*}[!htb]
    \centering
    \includegraphics[width=0.7\linewidth]{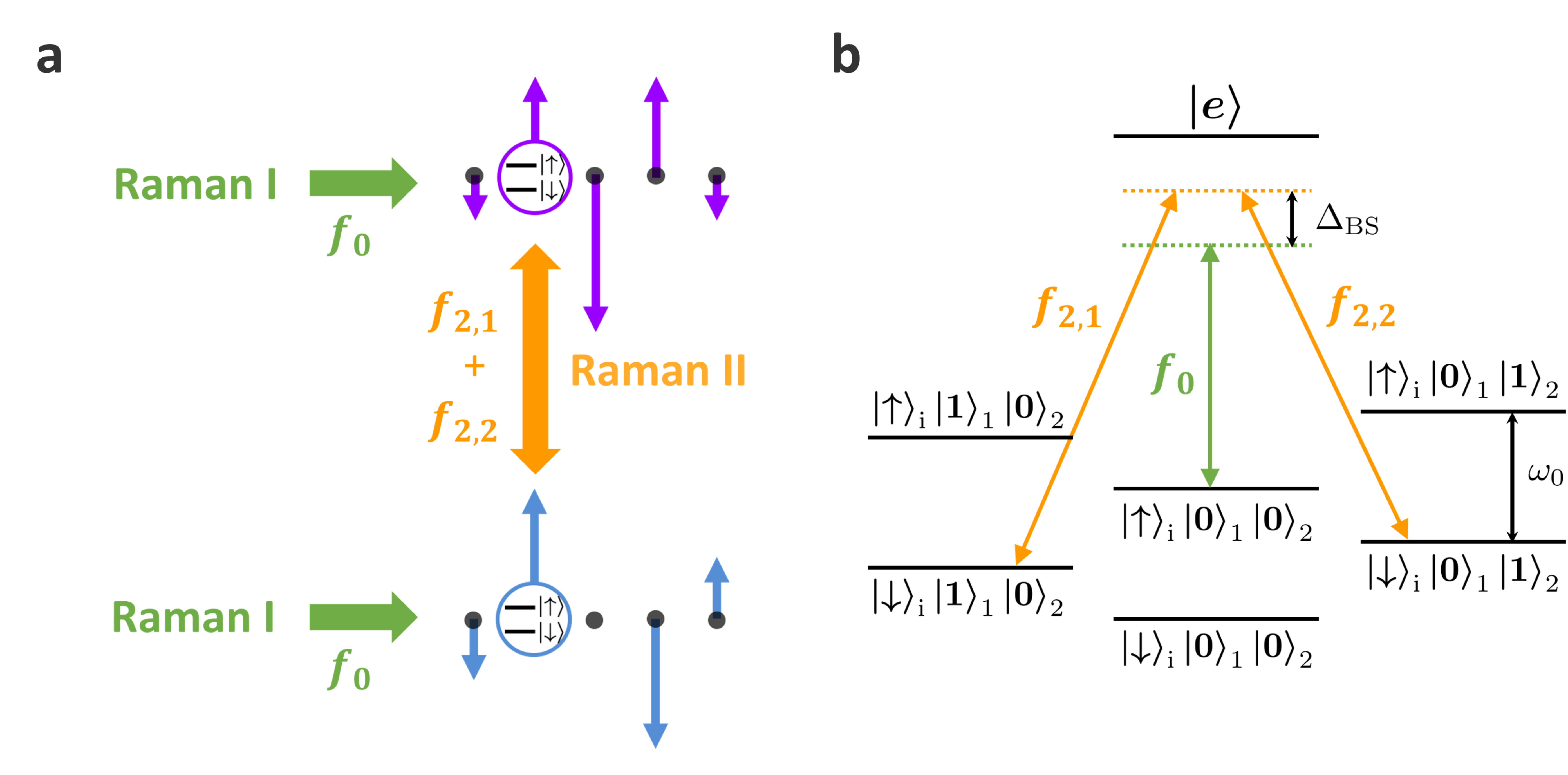}
    \caption{
    \textbf{Raman schematic diagram of beam-splitter}. $\bm{\textbf{a}}$, Frequency arrangement of two Raman lasers from perpendicular directions. Raman I is a global laser with one frequency component, and Raman II is an individually addressed laser with two components focused on one ion. $\bm{\textbf{b}}$, Energy diagram of beam-splitter. $\Delta_{\text{bs}}$ is a frequency detuning between Raman I and Raman II, effectively introducing two off-resonant RSB on a single ion. $\omega_0$ is the frequency of ion-qubit. Here two energy levels are connected by the Raman transition, $\left| \downarrow\right\rangle_i \left|1\right\rangle_1\left|0\right\rangle_2$ and $\left| \downarrow\right\rangle_i\left|0\right\rangle_1\left|1\right\rangle_2$.
    } 
    \label{fig:Figure_C1}
\end{figure*}

An example of detailed beam-splitter scheme is illustrated in Fig.~\ref{fig:Figure_C1} for mode 1 and mode 2 with ion 2. We simultaneously apply a pair of off-resonant sideband transitions with a detuning of $\Delta_{\text{bs}}$ on ion 2. One of the Raman lasers has a frequency of $f_0$, and the other has two frequencies of $f_{2,1}$ and $f_{2,2}$, where $f_{i,m}$ denotes the frequency applied to ion $i$ and couples to mode $m$, as shown in Fig.~\ref{fig:Figure_C1}a. Fig.~\ref{fig:Figure_C1}b is the energy diagram of this transition. We connect $\left|\downarrow\right\rangle_2 \left|1\right\rangle_1\left|0\right\rangle_2$ and $\left|\downarrow\right\rangle_2 \left|0\right\rangle_1\left|1\right\rangle_2$ to $\left|\uparrow\right\rangle_2 \left|0\right\rangle_1\left|0\right\rangle_2$ using off-resonant red sideband transitions of mode 1 and mode 2, respectively. In practice, we choose $f_{2,m}-f_0 = \Delta_{\text{bs}}+\omega_0-\nu_{m}$, where $\Delta_{\text{bs}}$ denotes the off-resonant detuning, $\omega_0$ the energy gap of the ion-qubit and $\nu_m$ the frequency of mode $m$. Ignoring all the off-resonant terms, the effective Hamiltonian of this system is written as
\begin{equation}
\label{eq:beam-splitter}
H_{\text{bs}^{(1,2)},\text{eff}}=-\frac{\eta_{2,1}\eta_{2,2}\Omega_{2,1}\Omega_{2,2}}{4} \frac{1}{\Delta_{\text{bs}}}\sigma_{2,z}(a_1a_2^{\dagger}+a_1^{\dagger}a_2)+H_{\text{ac}},
\end{equation}
where
\begin{equation}
\label{eq:ac Starkshift}
\begin{aligned}
H_{\text{ac}}^{(1,2)}=&\text{ }
\frac{\eta_{2,1}^2}{4}(\frac{\Omega_{2,1}^2}{\Delta_{\text{bs}}}-\frac{\Omega_{2,2}^2}{\nu_2-\nu_1-\Delta_{\text{bs}}})a_1a_1^{\dagger}\sigma_{2,z}\\
+&\text{ }\frac{\eta_{2,2}^2}{4}(\frac{\Omega_{2,2}^2}{\Delta_{\text{bs}}}-\frac{\Omega_{2,1}^2}{\nu_1-\nu_2-\Delta_{\text{bs}}})a_2a_2^{\dagger}\sigma_{2,z}\\
=&\text{ }\omega_1^{\prime}a_1a_1^{\dagger}\sigma_{z}+\omega_2^{\prime}a_2a_2^{\dagger}\sigma_{z}.
\end{aligned}
\end{equation}
Eq.~(\ref{eq:beam-splitter}) shows a spin-dependent beam-splitter between two vibrational modes with an effective Rabi frequency of $\eta_{2,1}\eta_{2,2}\Omega_{2,1}\Omega_{2,2}/\left(4\Delta_{\text{bs}}\right)$, where $\eta_{2,m}$ and $\Omega_{2,m}$ are the Lamb-Dicke parameter and the coupling strength between mode $m$ and ion 2, respectively. Eq.~(\ref{eq:ac Starkshift}) is the ac Stark shift term. We can compensate for the ac Stark effect by adding a corresponding frequency shift on $f_{2,1}$ as $f_{2,1}^{\prime}=f_{2,1}+\sigma_{2,z}(\omega_1^{\prime}-\omega_2^{\prime})$. The amount of the ac Stark shift is around several hundred Hz and we note that compensation of the ac Stark shift is important for the proper operation of beam-splitters. The experimentally-measured ac Stark shifts are in agreement with the theoretical values. We note that we can simultaneously perform two or more beam-splitters by applying properly-selected multiple-frequencies on Raman II with careful compensation of ac Stark shifts. We can use different ions for different pairs of modes, for example, ion 1 for mode 1 and 2 and ion 2 for mode 3 and 4, etc to reduce the cross-talk when simultaneous beam-splitters are applied.

\subsubsection{Fidelity and duration of beam-splitters}
The fidelity and duration of the beam-splitter are mainly determined by the choices of both $\Delta_{\text{bs}}$ and $\Omega_j$, since major errors come from the off-resonant coupling to other sideband transitions and carriers. Fig.~\ref{fig:Figure_C2} shows the numerical simulation with an average mode spacing of $\delta\bar{\nu}_{\text{M}}=0.05$ MHz. The fidelity and duration of the beam-splitter increase with $R_1=\Delta_{\text{bs}}/\sqrt{\eta_{j,m}\Omega_{j,m}\eta_{j,n}\Omega_{j,n}}$ and $R_2=\delta\bar{\nu}_{\text{M}}/\Delta_{\text{bs}}$, which means a long duration of the beam-splitter can suppress infidelities from the off-resonant coupling. In the experiment, we set $R_1$ from 1.5 to 3 and $R_2$ from 3 to 7. We also use a pulse-shaping method by modulating Omega with a sinusoidal function at the beginning and the end of the pulse.

Fig.~\ref{fig:Figure_C3} shows the numerical simulation of the beam-splitters with systematic errors, which is performed by solving the Lindblad master equation with Lindblad operators as $\hat{L}_{\text{heating}} = \alpha_m a_m^{\dagger}a_m$, $\hat{L}_{\text{mdc}} = \sum_{m}\sqrt{\kappa_m} a_m^{\dagger}a_m$, and $\hat{L}_{\text{sdc}} = \sqrt{\kappa_i} \sigma_{z,i}$. Here $\alpha_m$ and $\kappa_m$ denote the heating and decoherence rates of mode $m$, $\kappa_i$ the decoherence rate of ion $i$. We perform the simulation with an average mode spacing of 50 kHz and a 50:50 beam-splitter duration of 250 $\mu s$. According to the fitting curves, the error is proportional to the heating rate and inversely proportional to the coherence time in the interval where the coherence time is significantly longer than that of the beam splitter. Green stars denote the measured system heating rate (See Method B) and coherence times.

\subsubsection{Fidelity measurements of the beam splitters}
\begin{figure*}[!htb]
    \centering
    \includegraphics[width=0.8\linewidth]{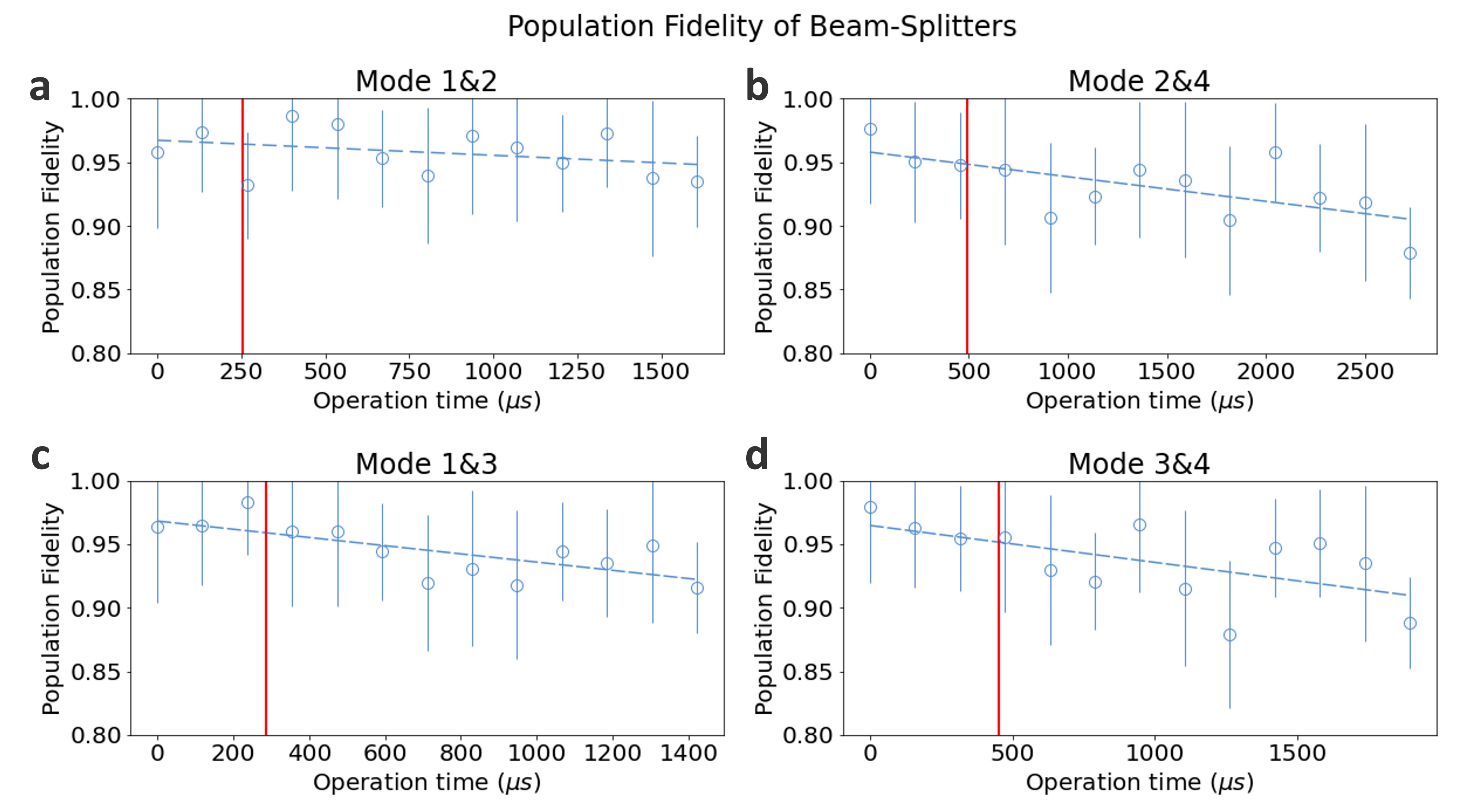}
    \caption{
    \textbf{Fidelity measurements of beam-splitters}. We calculate the population fidelity of beam-splitters between modes ($\bm{\textbf{a}}$) 1 and 2, ($\bm{\textbf{a}}$) 2 and 4, ($\bm{\textbf{a}}$) 1 and 3, ($\bm{\textbf{a}}$) 3 and 4, using datas shown in Fig.~\ref{fig:Figure_2}. The blue circles represent the calculated fidelity, and the blue dashed lines represent the linear fitting results. The intersection of the blue and red lines represents the fidelities of the 50:50 beam splitters.
    } 
    \label{fig:Figure_C4}
\end{figure*}

We verify the performance of the beam splitters by comparing the probability distributions shown in Fig.\ref{fig:Figure_2} with the ideal probability distribution. We estimate the population fidelity of the beam splitter between mode $m$ and $n$ by using the formula of $F^{(m,n)} = \left|\sum_{k} \sqrt{ P_{k,\text {ideal}}^{(m,n)} P_{k,\text{exp}}^{(m,n)} }\right|^{2}$, where $k$ denotes different output states, $P_{k,\text {ideal}}^{(m,n)}$ is the ideal output probability of the state $\ket{k}$, and $P_{k,\text{exp}}^{(m,n)}$ is the measured output probability of the state $\ket{k}$ in the experiment as shown in Fig.~\ref{fig:Figure_2}. We estimate the fidelity of the beam splitter through a linear fitting model $F^{(m,n)}_{\textrm{fit}}(t) = F^{(m,n)}_{\textrm{ini}} + \epsilon^{(m,n)} t$, where $F^{(m,n)}_{\textrm{ini}}$ denotes the initial population fidelity which is mainly limited by imperfect state preparation and detection, $\epsilon^{(m,n)}$ is a fitting parameter for decay of fidelities from errors caused by off-resonant coupling, heating and decoherence. The measured probability fidelities and the fitting results are shown in Fig.~\ref{fig:Figure_C4}. Here the average value of $F^{(m,n)}_{\textrm{ini}}$ for four different beamsplitters is $96.46\%$ and the average fidelity at the time of the beamsplitter  $F^{(m,n)}(t_{\textrm{bs}}^{(m,n)})$ is $95.58\%$, where $t_{\textrm{bs}}^{(m,n)}$ denotes the duration of 50:50 beam splitter between mode $m$ and $n$. The result reveals that the error induced by a 50:50 beam-splitter itself is less than $1\%$.

\subsection{Compensation of phase shift induced by ac Stark shift}

The laser components for a beam-splitter between two modes cause ac Stark shifts for all the other modes. The amount of the phase shift for the $k$-th mode introduced by the ac Stark shifts from a beam-splitter between mode $m$ and $n$ is written as

\begin{equation}
\int_{T_{i}}^{T_{f}}\frac{(\eta_{j,k}\Omega_{j,m}(t))^2}{2\delta_{j,k,m}}+\frac{(\eta_{{j,k}}\Omega_{{j,n}}(t))^2}{2\delta_{j,k,n}}dt,
\end{equation}
where $T_i$ and $T_f$ are the starting and ending times of the beam-splitter, and $\delta_{j,k,m} = (f_{j,m}-f_0) -(\omega_0-\nu_{k})$ is the detuning of the laser component to the RSB of $k$-th mode, respectively. The Rabi frequency is time-dependent because of the pulse shaping. The beam-splitter performance is shown in Fig.~\ref{fig:Figure_2}, and the related experimental parameters and fidelities are listed in Table.~\ref{tab:bsperformance}. In the interferometric configuration, the phase shift of each beam-splitter is calculated by including the effects of ac Stark shifts from all the previous beam-splitters, which is consistent with the experimental shift.

\subsection{Interferometric configurations for tomography}

The unitary rotation matrix of the interferometric configurations is labeled as $U(g)$, where the number $g$ denotes different settings. Then a superoperator~\cite{banchi2018multiphoton} can be constructed by $\mathcal{L}_{\nu g,\alpha\beta}=\langle \nu | U(g)^{\dagger}| \alpha \rangle \langle \beta | U(g) | \nu \rangle$, where the number $\alpha, \beta$ denotes the elements of the Fock space for the input states and $\nu$ for the output states. Then we get the best choice for tomography when $\det (\mathcal{L}^{\dagger} \mathcal{L})$ has the largest value. Based on the output probabilities, the reconstruction of the input density matrix is realized by $\rho_{\textrm{best }}:=\left(\mathcal{L}^{\dagger} \mathcal{L}\right)^{-1} \mathcal{L}^{\dagger}[p]$~\cite{banchi2018multiphoton}, where $p$ is the output probabilities. Due to errors in the experiment, a maximum-likelihood method is used to predict the possible reconstructed density matrix. We assume a positive density matrix $\rho_{\textrm{best}}^{\prime}$ with $Tr[\rho_{\textrm{best}}]=1$. By minimizing the 2-norm $\lvert \rho_{\textrm{best}}^{\prime} - \rho_{\textrm{best}}\rvert$, we get the possible density matrix for our reconstruction scheme.

For a system with the input state of two-mode and one-phonon, we need three interferometric configurations~\cite{banchi2018multiphoton}, where each consists of one beam-splitter. In the experiment, we choose the rotating phases of the three beam-splitters as $\{0,2/3 \pi,4/3 \pi\}$ with an optimized rotating angle of 0.304$\pi$. Table.~\ref{tab:tworesult} shows the reconstructed matrix of various one-phonon states of two transverse modes of three ions.

The number of interferometric configurations can be reduced to one with two additional vacuum modes. Table.~\ref{tab:fourmodeset} shows an interferometric configuration with four different beam-splitters. This configuration can be used to reconstruct any phonon-number state between two modes~\cite{banchi2018multiphoton}. The reconstruction results are shown in Table.~\ref{tab:fourresult}.

\subsection{Scaling up of the network}

The phononic networks can be scaled up to over hundred modes which will be enough to demonstrate quantum advantage \cite{zhong2020quantum,chen2021quantum}. We address the problems in scaling up the phononic network such as the decrease of mode-frequency separation, performance degrade of the beam splitters, and coherence time of the modes below in detail. 
\subsubsection{Mode-frequency separation} 

We can increase the number of modes by increasing the number of ions $N$ in the trap. This results in decreasing mode-frequency separation. For the transverse mode, which is the mode of interest in our phononic network, the COM mode has the largest frequency and the other modes are distributed with smaller frequencies, where the smallest frequency is mainly determined by the ratio of axial and radial COM mode frequencies. For example, when the axial-COM frequency is large enough to produce zig-zag structure of ions, the smallest frequency of the transverse mode gets close to zero. The distribution of the transverse-mode frequency is mainly determined by distances between ions. Roughly speaking, when ions are equally-spaced, the distribution of mode frequencies are close to uniform. Therefore, with the equal-spacing of linear-ion chain, the average separation of transverse-mode-frequencies can be estimated as $\delta\bar{\nu}_{\text{M}}=(\nu_{\rm COM}-\nu_{\rm min})/N$, where $\nu_{\rm COM}$ is the frequency of the transverse COM mode, $\nu_{\rm min}$ is the smallest frequency of the transverse mode, and $N$ is the number of ions and modes. For typical parameters of $\nu_{\rm COM}= 5$ MHz, $\nu_{\rm min}= 1$ with $N=100$, the average of mode-frequency separation can be around 40 kHZ.

In the case of $^{171}$Yb$^+$ ions, the estimated frequency separation can be considered as the upper bound. It is because the spacing of ions needs to be 1.5 $\mu$m, which is challenging to individually address ions in experiment, to achieve the mode-frequency separation of 40 kHZ with $^{171}$Yb$^+$ ions. We note that in the case of $^{9}$Be$^+$ ions, the spacing can be as large as 4.1 $\mu$m, where the individual addressing can be realized without serious difficulty. Instead, we estimated the mode-frequency separation with the reasonable spacing of $^{171}$Yb$^+$ ions as 5 $\mu$m and 3.5 $\mu$m, which reduces to the order of a few kHz for one hundred ions as shown in Fig.~\ref{fig:Figure_F}(a).

\subsubsection{Performance of beamsplitters with large number of modes}

We discuss the performance of beamsplitters as the increase of the number of ions in terms of fidelity and connectivity as follows.

\begin{figure*}[!htb]
    \centering
    \includegraphics[width=0.9\linewidth]{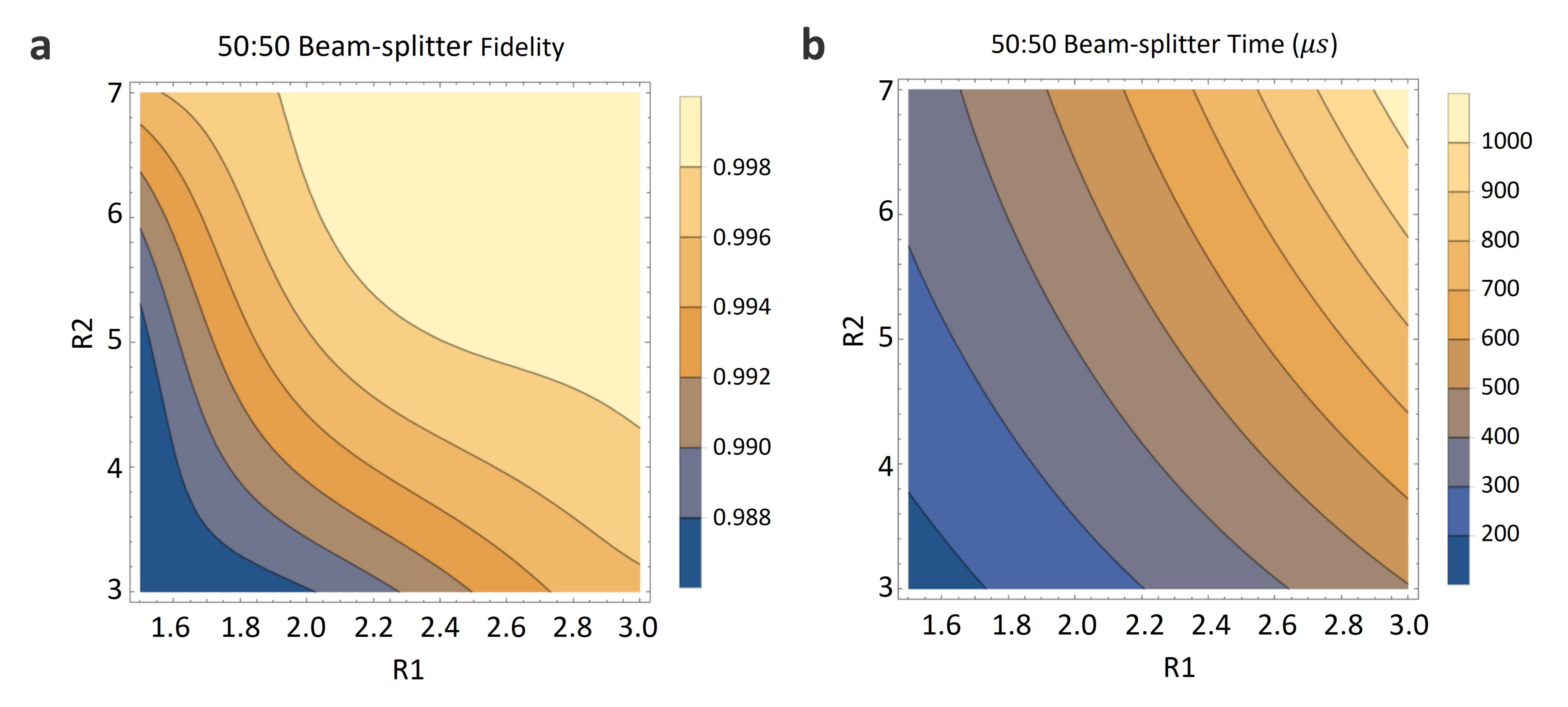}
    \caption{
    \textbf{Numerical simulation for fidelity and duration of 50:50 beam-splitters}. $\bm{\textbf{a}}$, Fidelity of a 50:50 beam-splitter with  $R_1=\Delta_{\text{bs}}/\sqrt{\eta_{j,m}\Omega_{j,m}\eta_{j,n}\Omega_{j,n}}$ and $R_2=\delta\bar{\nu}_{\text{M}}/\Delta_{\text{bs}}$, where $\eta_{j,m}$ and $\eta_{j,n}$ are Lamb-Dicke parameters of two modes and $\Omega_{j,m}$ is the Rabi frequency of frequency component $f_{j,m}-f_{0}$. $\bm{\textbf{b}}$, Prediction of operation time for a 50:50 beam-splitter. Here the simulation based on an average mode spacing of $50$kHz.
    } 
    \label{fig:Figure_C2}
\end{figure*}

\begin{figure*}[!htb]
    \centering
    \includegraphics[width=1.0\linewidth]{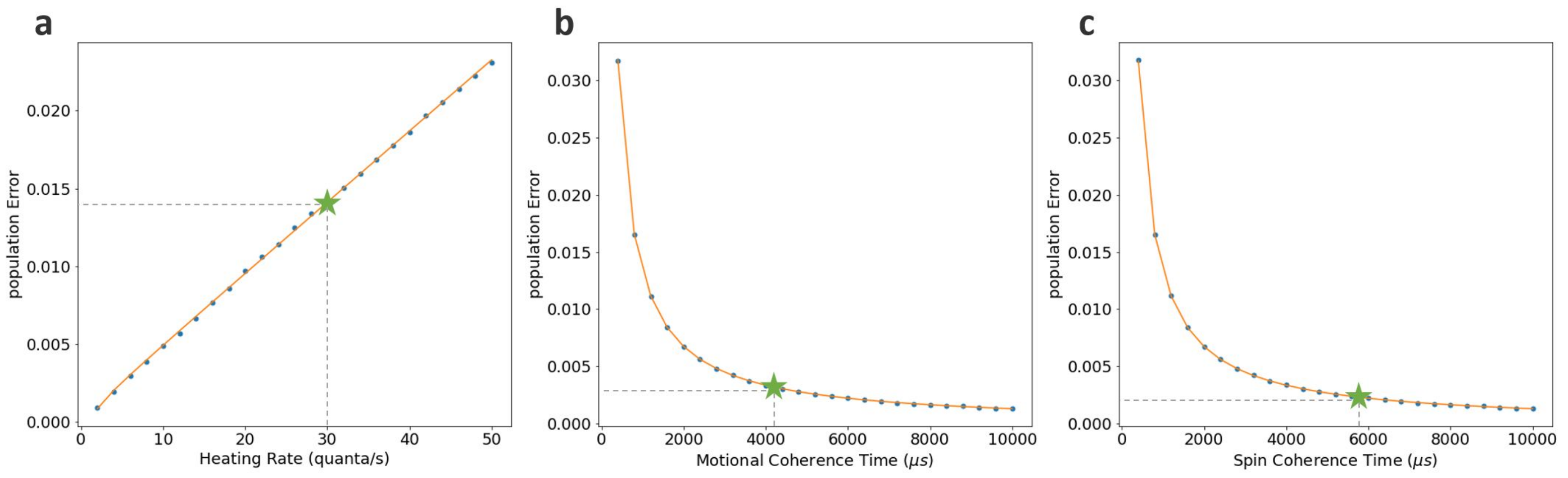}
    \caption{
    \textbf{Numerical simulation for systematic errors of beam-splitters}. We simulate the performance of 50:50 beam-splitters with similar parameters used in the experiment. We achieve population errors induced by ($\bm{\textbf{a}}$) heating, ($\bm{\textbf{b}}$) motional decoherence, and ($\bm{\textbf{c}}$) spin decoherence by subtracting inevitable errors caused by off-resonant couplings shown in Fig.~\ref{fig:Figure_C2}a. The blue points represent simulation results, with the orange fitting curves representing the error trend. The green stars represent the measurement results of the corresponding values. 
    } 
    \label{fig:Figure_C3}
\end{figure*}

{\bf Fidelity:} Typically, the decrease of mode-frequency separation can introduce the decrease of fidelities of beamsplitter by Raman transitions due to additional off-resonant coupling to spectator modes. The problem of infidelity of beam-splitter can be resolved by the decrease of strength of the beamsplitter with the scale of $1/N$, that is, the increase of the beamsplitter duration with the number of ions $N$. It is because that as shown in Fig.~\ref{fig:Figure_C2}, the fidelities of the beam-splitters can be maintained by keeping the ratios of $R_1=\Delta_{\text{bs}}/\sqrt{\eta_{j,m}\Omega_{j,m}\eta_{j,n}\Omega_{j,n}}$ and $R_2=\delta\bar{\nu}_{\text{M}}/\Delta_{\text{bs}}$. As discussed in the previous section, the average separation of transverse-mode-frequency $\delta\bar{\nu}_{\text{M}}$ (and $\Delta_{\text{bs}}$) and the mode coupling strength $\eta$ scales as $1/N$ and $1/\sqrt{N}$ with the number of ions $N$, respectively. In order to keep the ratio of $R_1$ and $R_2$, therefore, we need to reduce  $\Omega_{j,m}$ and $\Omega_{j,n}$ with the scale of $1/\sqrt{N}$. This results in the increase of the beamsplitter duration proportionally to the number of ions $N$, since the effective Rabi frequency of the beamsplitter $\eta_{j,m}\Omega_{j,m}\eta_{j,n}\Omega_{j,n}/(4\Delta_{\text{bs}})$ as shown in Eq.~\ref{eq:beam-splitter}. The durations of beamsplitter depending on the number of ions to maintain the fidelity are shown in Fig.~\ref{fig:Figure_F}(b). Similarly, all the sideband operations used for phonon state preparation and detection should scale as the same level to avoid additional errors from off-resonant coupling.

{\bf Connectivity:} In the phononic network, beamsplitters are performed by using an ion that has the large coupling strengths for the two modes of interest. Here we show that we can find proper ions that couple all pairs of modes with sufficient couple strengths with large number of ions. To verify the connectivity of large-scale phononic system, we numerically search the parameters up to 100 modes of 100 ions and found that 99.9 $\%$ of the largest product of coupling strengths in all the pairs are greater than $\eta^2/N$, which is the square of $\eta/\sqrt{N}$, the coupling strength of the COM mode. Here, the average and standard deviations of largest product of coupling strengths are 1.95$\eta^2/N$ and 0.12 $\eta^2/N$, respectively.

\subsubsection{Coherence time and heating rate of the large scale phononic network}
\begin{figure*}[!htb]
    \centering
    \includegraphics[width=1\linewidth]{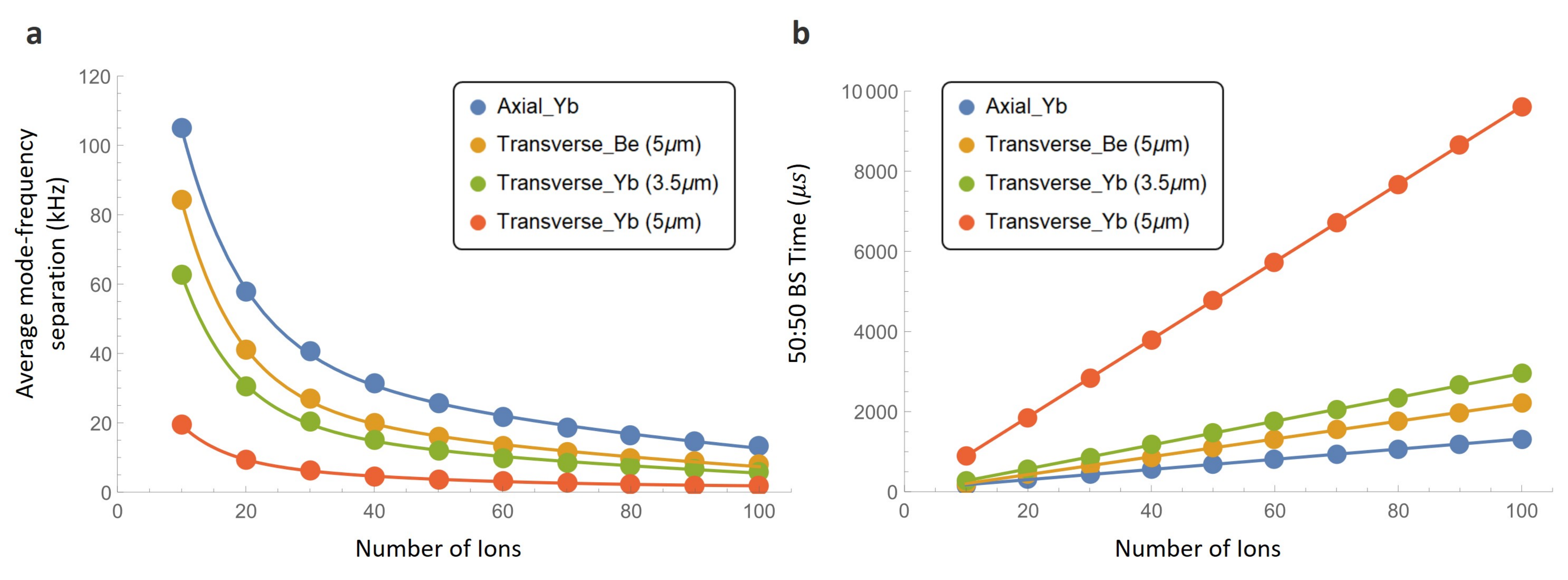}
    \caption{
    \textbf{Average mode frequency spacing and operation time with the number of ions}. $\bm{\textbf{a}}$, Numerical simulation of mode frequency spacing with an increasing number of ions. $\bm{\textbf{b}}$, Average 50:50 beam-splitter operation time based on the results as shown in panel a. Here the choice of $R_1$ and $R_2$ keeps a fidelity of about $99.5\%$ for the beam-splitter.
    } 
    \label{fig:Figure_F}
\end{figure*}

Coherence time and the duration of heating of the network should be larger than the total duration of the interferometer. A fully connected interferometer with $N$ modes requires an order of $N^2/2$ number of beam-splitters. Assuming we can perform $N/2$ beamsplitters simultaneously as discussed in Method C, the total duration of the fully connected interferometer can be $N$ times the duration of a single beam splitter. It would be several milliseconds for the case of one hundred $^{171}$Yb$^+$ ions with the spacing of 5 $\mu$m as shown in Fig.~\ref{fig:Figure_F}(b).

The coherence time of the transverse mode itself is mainly limited by the fluctuation of RF power. But, we found that the coherence time of the transverse-mode interferometer is related to the stability of DC voltages, not the RF power. It is because the coherence time of the beamsplitter is determined by the stability of frequency difference between the relevant two modes, not each mode frequency. In the linear Paul traps, DC-voltages are applied for the confinement of ions along the axial direction, which determines the frequency separation of transverse modes. The stabilization technology for the DC-voltage is well-developed. In order to reach a coherence time of 1 second for the vibration mode with 1 MHz, the voltage stability should be in the order of 10$^{-6}$, which is much less than state-of-the-art voltage sources at the 10$^{-11}$ stability level~\cite{rufenacht2018impact}.

\begin{figure*}[!htb]
    \centering
    \includegraphics[width=1\linewidth]{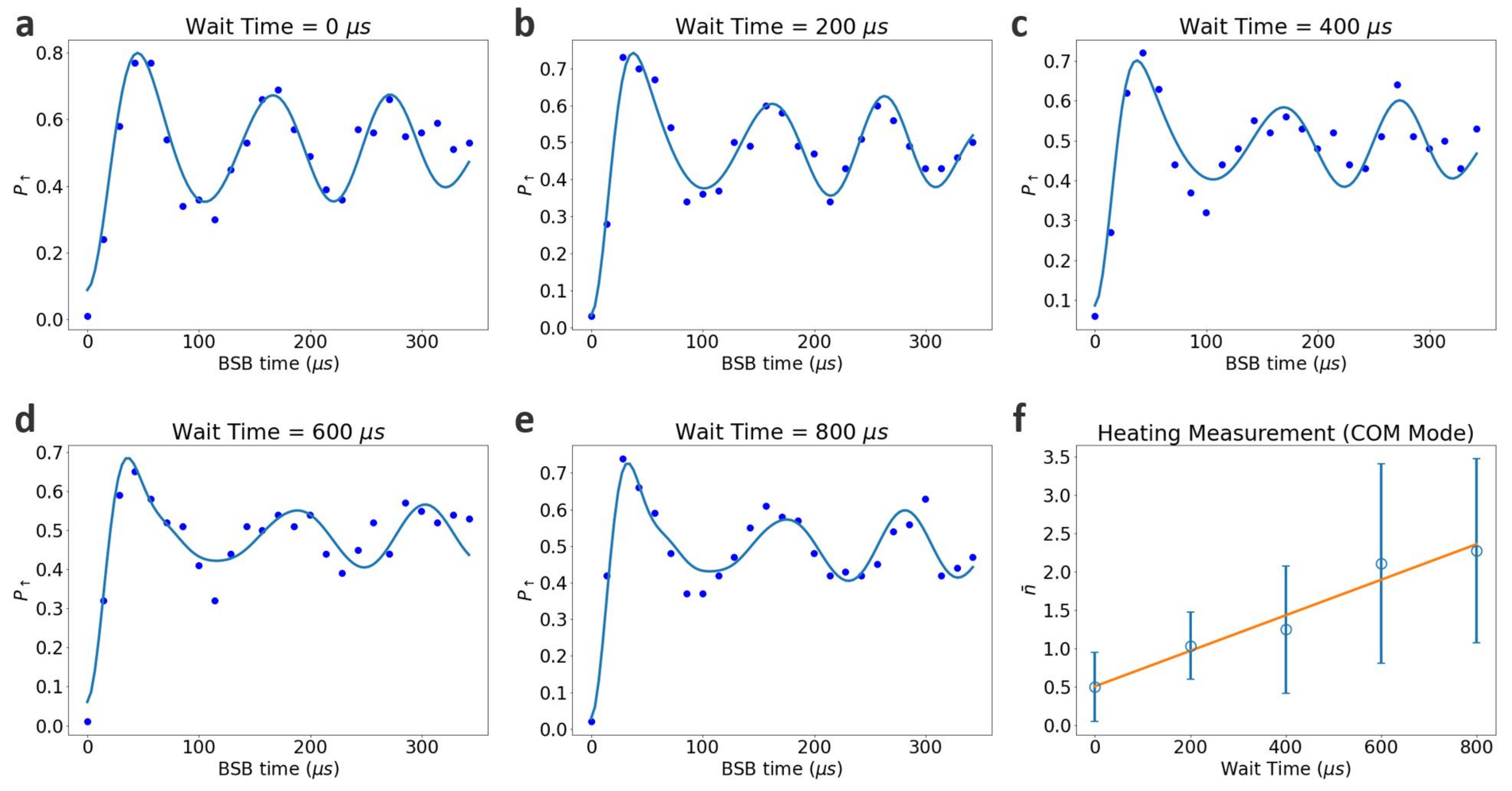}
    \caption{
    \textbf{Heating measurements of COM mode in transverse direction.} \textbf{a}-\textbf{e}, time evolutions of blue sideband (BSB) transitions after different waiting times. The blue dots are the measured probabilities of the spin-up state, while the curves are fitting results based on different thermal state distributions. \textbf{f}, average phonon numbers after different waiting times. We estimate a heating rate of $2.3 \times 10^3$ quanta/s.
    } 
    \label{fig:Figure_B1}
\end{figure*}

\begin{figure*}[!htb]
    \centering
    \includegraphics[width=0.72\linewidth]{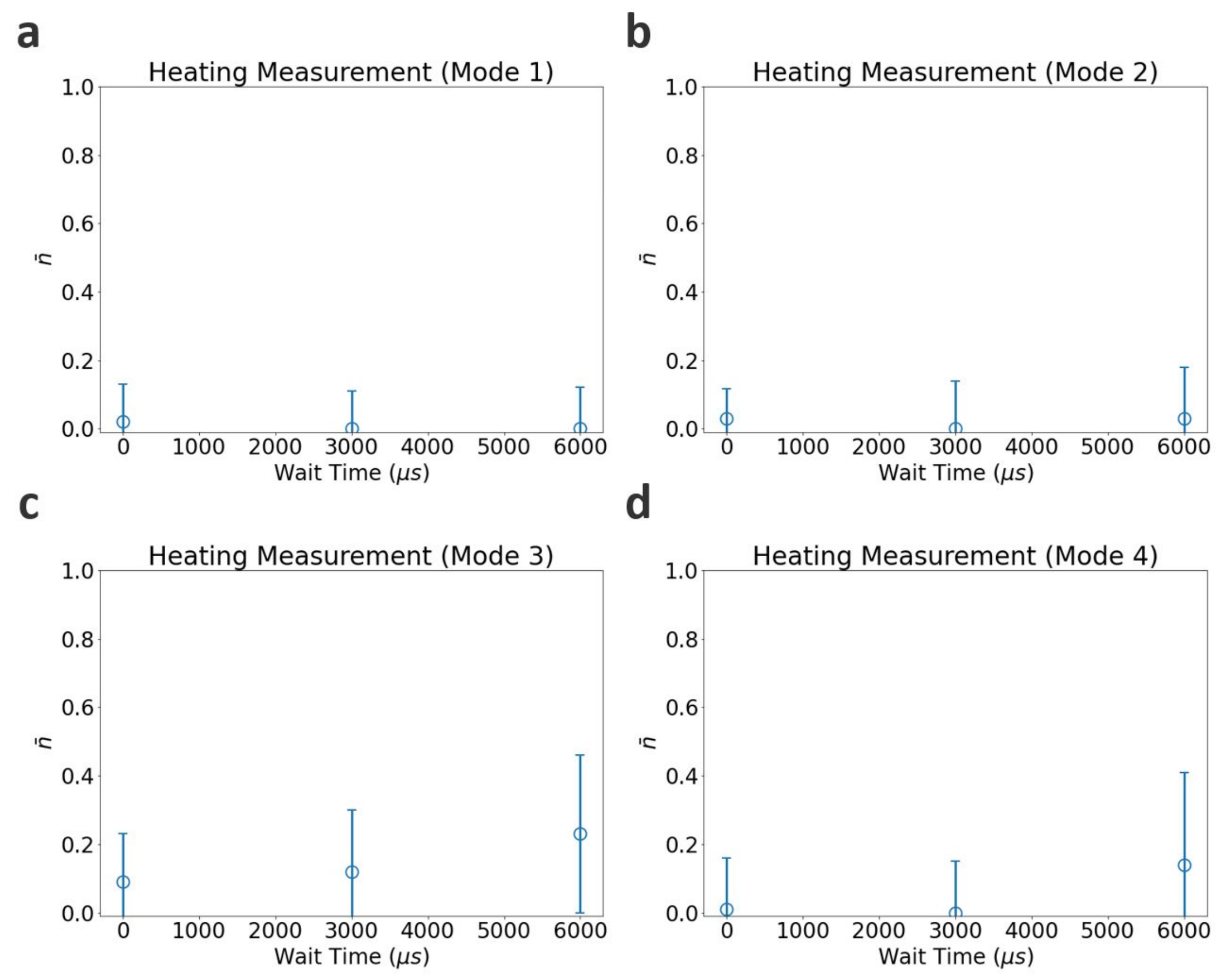}
    \caption{
    \textbf{Heating measurements of other transverse modes.} \textbf{a}-\textbf{d}, Average phonon numbers on mode 1 to mode 4 after various waiting times up to 6000 $\mu s$. We obtain the average phonon number $\bar{n}$ by fitting the time evolution of the BSB with the thermal state distribution.
    }
    \label{fig:Figure_B2}
\end{figure*}

In experiment, the coherence time of 0.2 second has been reported for the axial mode of a single ion system~\cite{lucas2007long}, where the limitation was not from the fluctuation of DC voltage, but the heating of the mode. In the system with large number of ions, the heating rate of the COM modes proportionally increases with the number of ions~\cite{lechner2016electromagnetically}. However, the heating of the other collective modes are orders of magnitude suppressed which was observed in our experiment (Fig.~\ref{fig:Figure_B1} and Fig.~\ref{fig:Figure_B2}) and in Ref.~\cite{kalincev2021motional}. Authors in Ref.\cite{kalincev2021motional} reported no significant heating of any of next-to-COM transverse modes with below 1 MHz mode-frequency and 10 - 100 $\mu$m sizes of ion chains. Other than COM modes are less heated because the electrical noise in the system are not localized below the size of ion chains. For systems with over hundred ions, the size of the chain can be order of millimeter and some of collective modes can be heated. As we did not include the COM mode in the experiment, we also can exclude some of high heating modes in that case. In the extreme of localized noise, the heating rates of other-than-COM collective modes approach to that of single ion~\cite{brownnutt2015ion}. In principle, the heating rates can be further suppressed by locating the trap in cryo-environment~\cite{pagano2018cryogenic}, which may suppress the heating rates by two to four orders of magnitude~\cite{deslauriers2006scaling,brownnutt2015ion}.
 
Finally, as shown in Fig.~\ref{fig:Figure_F}, alternative trapped ion systems can also be considered for a shorter operating duration. For example, if we reduce the spacing of $^{171}$Yb$^+$ ions to 3.5 $\mu$m, the mode spacing increases three times, and the duration reduces by the same amount. The duration can be further reduced using axial vibrational modes. For the ion spacing of 5$\mu m$, the axial-mode-frequency separation approaches 20 kHz, and the duration of the beam splitters becomes below 1 ms for one hundred modes. However, for the axial modes, many of low-frequency-modes roughly below 1 MHz need to be excluded in the interferometer. We note that it can be considered to use a lighter ion such as $^9$Be$^+$, which reduce the duration by square-root of mass ratio, roughly 4 times from the same condition of $^{171}$Yb$^+$ ions as shown in Fig.~\ref{fig:Figure_F}. All these simulations are performed with the fidelity of the beam-splitter of $99.5\%$. In short, we do not think there exist fundamental or technological bottle neck to build a large-scale phononic network including over hundred vibrational modes.

\begin{table*}[!htb]
\caption{\label{tab:bsperformance}%
Parameters and fidelities of 50:50 beam-splitters in four-mode setup. Here the beam-splitter fidelity is measured by the population overlap of the experimental state and the ideal state, where errors are mainly from the imperfections of state preparation and detection (SPD).
}
\begin{ruledtabular}
\begin{tabular}{ccccccc}
\textrm{Ion}&
\textrm{Modes ($m_1\&m_2$)}&
\textrm{$\Delta_{\text{bs}}$(kHz)}&
\textrm{$\eta_{m_1} \Omega_1$(kHz)}&
\textrm{$\eta_{m_2} \Omega_2$(kHz)}&
\textrm{Duration($\mu s$)}&
\textrm{Fidelity (With SPD Errors)}\\
\colrule
3 & $1\&3$ & -10 & -6.3 & -4.4 & 286.6 & 95.89 $\pm$ 1.26$\%$\\
4 & $2\&4$ & -10 & -6.3 &  3.1 & 468.1 & 94.85 $\pm$ 2.70$\%$\\
5 & $3\&4$ & -10 &  4.4 &  3.1 & 453.7 & 95.15 $\pm$ 1.19$\%$ \\
2 & $1\&2$ & -10 &  6.3 &  6.3 & 254.8 & 96.43 $\pm$ 1.26$\%$ \\

\end{tabular}
\end{ruledtabular}
\end{table*}

\begin{table*}[!htb]
\caption{\label{tab:fourmodeset}
Interferometric configuration for a two-mode input state with two ancillary modes. Here only one configuration is used. The order keeps the same as in the real experiment.
}
\begin{ruledtabular}
\begin{tabular}{cccc}
&\textrm{Modes of beam-splitter}&
\textrm{$\{\theta,\phi\}$}&\\
\colrule
&$1\&3$ & $\{0.696\pi,0\pi\}$&\\
&$2\&4$ & $\{0.304\pi,0\pi\}$&\\
&$3\&4$ & $\{0.5\pi,0.5\pi\}$&\\
&$1\&2$ & $\{0.5\pi,1.0\pi\}$&\\
\end{tabular}
\end{ruledtabular}
\end{table*}

\begin{table*}[!htb]
\caption{\label{tab:tworesult}Reconstruction results for two modes input states without using ancillary modes.}
\begin{ruledtabular}
\begin{tabular}{cccc}
 Input state & Ideal density matrix& Reconstructed matrix & Fidelity\\ \hline
 $|1\rangle_{1}|0\rangle_{2}$ & $\left(\begin{array}{cc}1 &0 \\ 0 & 0 \end{array}\right)$ & $\left(\begin{array}{cc}0.965 & 0.035+0.007 i\\ 0.035-0.007 i & 0.001 \end{array}\right)$ & $96.52 \pm 1.35\%$\\
 \\
 $|0\rangle_{1}|1\rangle_{2}$ & $\left(\begin{array}{cc}0 &0 \\ 0 & 1 \end{array}\right)$ & $\left(\begin{array}{cc}0 & 0.003+0.016 i\\ 0.003-0.016 i & 0.987 \end{array}\right)$ & $98.67 \pm 1.39\%$\\
 \\
 $\frac{1}{\sqrt{2}}(|1\rangle_{1}|0\rangle_{2}+|0\rangle_{1}|1\rangle_{2})$ & $\left(\begin{array}{cc}0.5 &0.5 \\ 0.5 & 0.5 \end{array}\right)$ & $\left(\begin{array}{cc}0.481 & 0.490+0.011 i\\ 0.490-0.011 i & 0.5 \end{array}\right)$ & $98.09 \pm 0.66\%$\\
 \\
 $\frac{1}{\sqrt{2}}(|1\rangle_{1}|0\rangle_{2}+i|0\rangle_{1}|1\rangle_{2})$ & $\left(\begin{array}{cc}0.5 &-0.5i \\ 0.5i & 0.5 \end{array}\right)$ & $\left(\begin{array}{cc}0.468 & 0.005-0.485 i\\ 0.005+0.485 i & 0.503 \end{array}\right)$ & $97.12 \pm 0.78\%$\\
 \\
 $\frac{1}{\sqrt{2}}(|1\rangle_{1}|0\rangle_{2}-i|0\rangle_{1}|1\rangle_{2})$ & $\left(\begin{array}{cc}0.5 &0.5i \\-0.5i & 0.5 \end{array}\right)$ & $\left(\begin{array}{cc}0.451 & -0.029+0.488 i\\ -0.029-0.488 i & 0.530 \end{array}\right)$ & $97.92 \pm 0.72\%$\\
 \\
 $\frac{1}{\sqrt{2}}(|1\rangle_{1}|0\rangle_{2}-|0\rangle_{1}|1\rangle_{2})$ & $\left(\begin{array}{cc}0.5 &-0.5 \\-0.5 & 0.5 \end{array}\right)$ & $\left(\begin{array}{cc}0.455 & -0.488+0.003 i\\ -0.488-0.003 i & 0.523 \end{array}\right)$ & $97.65 \pm 0.80\%$\\
 
\end{tabular}
\end{ruledtabular}
\end{table*}

\begin{table*}[!htb]
\caption{\label{tab:fourresult}Reconstruction results for two modes input states using two ancillary modes.}
\begin{ruledtabular}
\begin{tabular}{cccc}
 Input state & Ideal density matrix& Reconstructed matrix & Fidelity\\ \hline
 $\frac{1}{\sqrt{2}}(|1\rangle_{1}|0\rangle_{2}+|0\rangle_{1}|1\rangle_{2})$ & $\left(\begin{array}{cc}0.5 &0.5 \\ 0.5 & 0.5 \end{array}\right)$ & $\left(\begin{array}{cc}0.469 & 0.472-0.005 i\\
 \\ 0.472+0.005 i & 0.475 \end{array}\right)$ & $94.49 \pm 1.95\%$\\
 \\
 $\frac{1}{\sqrt{3}}(|1\rangle_{1}|1\rangle_{2}+i|0\rangle_{1}|2\rangle_{2}+i|2\rangle_{1}|0\rangle_{2})$ & $\left(\begin{array}{ccc}0.333 &0.333i &0.333i \\ -0.333i &0.333 &0.333 \\ -0.333i &0.333 &0.333 \end{array}\right)$ & $\left(\begin{array}{ccc}0.218 &0.063+0.297i &0.005+0.263i \\ 0.063-0.297i &0.423 &0.360+0.068i \\ 0.005-0.263i &0.360-0.068i &0.317 \end{array}\right)$ & $93.36 \pm 3.15\%$
\end{tabular}
\end{ruledtabular}
\end{table*}

\end{document}